\documentclass{amsart}
\usepackage[latin1]{inputenc}
\usepackage{amsmath,comment}
\usepackage{amsfonts,bbm}
\usepackage{amssymb}
\usepackage{graphicx}
\usepackage{graphics}
\usepackage{lscape}

\usepackage{amsmath}
\numberwithin{equation}{section}

\newtheorem{theorem}{Theorem}
\numberwithin{theorem}{section}
\newtheorem{lemma}{Lemma}
\numberwithin{lemma}{section}
\newtheorem{prop}{Proposition}
\numberwithin{prop}{section}
\newtheorem{corol}{Corollary}
\newtheorem{problem}{Problem}
\numberwithin{corol}{section}
\newtheorem{remark}{Remark}

\def\P{\mathbb{P}}

\def\R{\mathbb{R}}
\def\C{\mathbb{C}}
\def\E{\mathbb{E}}
\def\ind{\mathbbm{1}}

\title[Analysis of the SQF algorithm]{Stationary analysis of the "Shortest Queue First" service policy: the asymmetric case}

\author{Fabrice Guillemin}
\address{Orange Labs, CORE/TPN, 2 Avenue Pierre Marzin, 22300 Lannion}
\email{fabrice.guillemin@orange.com}

\author{Alain Simonian}
\address{Orange Labs, CORE/TPN, 38-40 Rue du G\'en\'eral Leclerc, 92794 Issy-les-Moulineaux}
\email{alain.simonian@orange.com}

\begin{document}

\begin{abstract}
As a follow-up to a recent paper considering two symmetric queues, the \textit{Shortest Queue First} service discipline is presently analysed for two general asymmetric queues. Using the results previously established and assuming exponentially distributed service times, the bivariate Laplace transform of workloads in each queue is shown to depend on the solution $\mathbf{M}$ to a two-dimensional functional equation
$$
\mathbf{M} = Q_1 \cdot \mathbf{M}\circ h_1 + Q_2 \cdot \mathbf{M}\circ h_2 + \mathbf{L}
$$
with given matrices $Q_1$, $Q_2$ and vector $\mathbf{L}$ and where functions $h_1$ and $h_2$ are defined each on some rational curve; solution $\mathbf{M}$ can then represented by a series expansion involving the semi-group $< h_1, h_2 >$ generated by these two functions. The empty queue probabilities along with the tail behaviour of the workload distribution at each queue are characterised.
\end{abstract}
\date{Version of \today}

\maketitle


\section{Introduction}


Given one server addressing two parallel queues $\sharp$1 and $\sharp$2, the  \textit{Shortest Queue First} (SQF) policy processes jobs according to the following rule. Let $U_1$ (resp. $U_2$) denote the workload in queue $\sharp$1 (resp. queue $\sharp$2) at a  given time, including the remaining amount of work of the job possibly in service; then
\begin{itemize} 
\item Queue $\sharp$1 (resp. queue $\sharp$2) is served if $U_1 \neq 0$, $U_2 \neq 0$ and $U_1 \leq U_2$ (resp. if $U_1 \neq 0, U_2 \neq 0$ and $U_2 < U_1$);
\item If only one of the queues is empty, the non empty queue is served;
\item If both queues are empty, the server remains idle until the next job arrival.
\end{itemize}
\noindent
The analysis of such a queuing discipline has been motivated in [\textbf{Gui12}] (and references therein), where general properties for the distribution of the pair $(U_1,U_2)$ are stated. Assuming Poisson job arrivals and generally distributed  service times, $(U_1,U_2)$ is a Markov process in $\mathbb{R}^+ \times \mathbb{R}^+$ whose stationary distribution is determined by its Laplace transform. The latter is then derived by solving the following 
\begin{problem}
\label{prob1}
Given some domain $\mathbf{\Omega} \subset \mathbb{C}^2$, determine analytic functions $G_1$, $G_2$ and $H$ in $\mathbf{\Omega}$ such that equations
$$
\left\{
\begin{array}{ll}
K_2(s_1,s_2)G_2(s_2) - H(s_1,s_2) = J_2(s_2) \; \; \; \mathrm{when} \; \; \; K_1(s_1,s_2) = 0,
\\ \\
K_1(s_1,s_2)G_1(s_1) + H(s_1,s_2) = J_1(s_1) \; \; \; \mathrm{when} \; \; \; K_2(s_1,s_2) = 0
\end{array} \right.
$$
together hold in $\mathbf{\Omega}$, where analytic functions $K_1$, $K_2$ and $J_1$, $J_2$ are given.
\end{problem}
\noindent
\indent
Assuming exponentially distributed service times, \textbf{Problem 1} has been solved in [\textbf{Gui12}] for the so-called \textbf{"symmetric case"}, that is, when arrival rates and service rates are the same for both queues. \textbf{Problem 1} then reduces to solving a single functional equation $M = q \cdot M \circ h + L$ for unknown function $M$, where given functions $q$, $L$ and $h$ are related to one branch of a cubic polynomial equation; the stationary distribution of any queue is then expressed by a series expansion involving all interated of function $h$. 

In the present paper, we intend to generalise the analysis to the so-called \textbf{"asymmetric case"} where arrival rates and service rates are generally distinct. As formulated in \textbf{Problem 1}, the curves defined in the $(O,s_1,s_2)$ plane by equations $K_1 = 0$ and $K_2 = 0$ are expected to play a central role. Specifically, we will show that curve $K_1 = 0$ (and \textit{mutatis mutandis}, curve $K_2 = 0$) verifies the following:
\begin{itemize}
\item it is a rational cubic (that is, with a rational parametrisation); 
\item there exists a rational mapping $\iota_1$ on cubic $K_1 = 0$ such that $\iota_1 \circ \iota_1 = \iota_1$ (that is, $\iota_1$ is an involution); 
\item when cut by a line $s_1 + s_2 = 2z$ with given $z > 0$, cubic $K_1 = 0$ defines 3 distinct intersection points $\mathfrak{A}_1$, $\mathfrak{B}_1$ and $\mathfrak{C}_1$; define $h_1(z)$ so that the transformed point $\iota_1(\mathfrak{A}_1)$ belongs to the line $s_1 + s_2 = 2h_1(z)$. The mapping $z \mapsto h_1(z)$ can then be extended as an analytic function on the complex plane cut along two distinct segments.
\end{itemize}
The above geometric and analytic properties will provide the key ingredients for the general resolution in the asymmetric case. 
\\
\indent
The paper is organised as follows. In Section \ref{GR}, main assumptions together with general results for functional equations and the analytic continuation of Laplace transforms are recalled for completeness from [\textbf{Gui12}]. In Section \ref{EXPI}, the general discussion is specified to the case of exponentially distributed service times. We then show in Section \ref{EXPII} that \textbf{Problem \ref{prob1}} above reduces to
\begin{problem} \textbf{(Asymmetric case)}
\label{prob2}
Solve the two-dimensional functional equation
$$
\mathbf{M}(z) = Q_1(z) \cdot \mathbf{M}\circ h_1(z) + Q_2(z) \cdot \mathbf{M}\circ h_2(z) + \mathbf{L}(z)
$$
for unknown vector function $z \mapsto \mathbf{M}(z) \in \mathbb{C}^2$, where matrices $Q_1$, $Q_2$ and vector $\mathbf{L}$ are given and functions $h_1$ and $h_2$ are defined as above. 
\end{problem}
\noindent
For real $z > 0$, the solution $\mathbf{M}(z)$ is written in terms of a series involving the semi-group $< h_1, h_2 >$ generated by functions $h_1$ and $h_2$ for the composititon operation; the empty queue probabilities, in particular, are expressed in terms of function $\mathbf{M}$ only. In Section \ref{AE}, function $\mathbf{M}$ is extended to some half-plane of the complex plane, enabling us (Section \ref{LQA}) to specify the tail behaviour of the workload distribution at each queue in relation to the associated preemptive HoL policy.


\section{General results}
\label{GR}


In this preliminary section, we recall the main assumptions and general results derived in [\textbf{Gui12}] for the SQF policy in the stationary regime. 

\subsection{Main assumptions}
\label{MA}

Incoming jobs are assumed consecutively enter queue $\sharp 1$ (resp. queue $\sharp 2$) according to a Poisson process with mean arrival rate $\lambda_1$ (resp. $\lambda_2$). Their respective service times are i.i.d. with probability distribution $\mathrm{d}B_1(x_1)$, $x_1 > 0$ (resp. $\mathrm{d}B_2(x_2)$, $x_2 > 0$) and mean  $1/\mu_1$ (resp. mean $1/\mu_2$). In the following, $\varrho_1 = \lambda_1/\mu_1$ (resp. $\varrho_2 = \lambda_2/\mu_2$) denotes the mean load of queue $\sharp 1$ (resp. queue $\sharp 2$) and we let $\varrho = \varrho_1 + \varrho_2$ denote the total load of the system. 
\\
\indent
Denote by $U_1(t)$ (resp. $U_2(t)$) the workload at queue $\sharp 1$ (resp. $\sharp 2$) at time $t$. With the above notation, the SQF policy then governs workloads $U_1(t)$ and $U_2(t)$ according to some evolution equations
which define the pair $\mathbf{U}_t = (U_1(t),U_2(t))$, $t \geq 0$, as a Markov process with state space $\mathcal{U} = \mathbb{R}^+ \times \mathbb{R}^+$ 
The distribution of process $\mathbf{U}$, in particular, does not give a positive probability to the diagonal $\{(u_1,u_1) \in \mathcal{U} \mid u_1 > 0\}$ in state space $\mathcal{U}$.


In the rest of the paper, the stability condition $\varrho = \varrho_1 + \varrho_2 < 1$ is assumed to hold, so that the stationary distribution of bivariate workload process $\mathbf{U}$ exists. Defining then the stationary distribution $\Phi$ of $\mathbf{U}$ by $\Phi(u_1,u_2)=\P(U_1\leq u_1,U_2 \leq u_2)$ for $u_1 \geq 0, u_2 \geq 0$, we assume that
\begin{itemize}
\item [\textbf{A.1}] $\Phi$ has a smooth density $\varphi_1(u_1,u_2)$ (resp. $\varphi_2(u_1,u_2)$) at any point $(u_1,u_2)$ such that $0 < u_1 < u_2$ (resp. $0 < u_2 < u_1$);
\item [\textbf{A.2}] $\Phi$ has a smooth density $\psi_1(u_1)$ (resp. $\psi_2(u_2)$) at any point $u_1$ (resp. $u_2$) on the boundary $\{(u_1,0) \mid u_1 > 0\}$ (resp. on the boundary $\{(0,u_2) \mid u_2 > 0\}$).
\end{itemize}
By assumptions \textbf{A.1}-\textbf{A.2} above, we can then write
\begin{align}
\mathrm{d}\Phi(u_1,u_2) = & \; (1-\varrho)\delta_{0,0}(u_1,u_2) \; + 
\label{dPhi}
\\ 
& \; \psi_1(u_1)\ind_{u_1 > 0}\mathrm{d}u_1 \otimes \delta_0(u_2) + \psi_2(u_2)\ind_{u_2 > 0}\mathrm{d}u_2 \otimes \delta_0(u_1) \; +
\nonumber \\
& \; (\varphi_1(u_1,u_2)\ind_{0 < u_1 < u_2} + \varphi_2(u_1,u_2)\ind_{0 < u_2 < u_1})\mathrm{d}u_1\mathrm{d}u_2
\nonumber
\end{align}
for all $(u_1,u_2) \in \mathcal{U}$.

\subsection{Functional equations}

Let $\mathbf{\Omega} = \{(s_1,s_2) \in \mathbb{C}^2 \mid \Re(s_1) > 0, \Re(s_2) > 0\}$ and its closure $\overline{\mathbf{\Omega}}$. The Laplace transforms $F_1$, $G_1$ of $\varphi_1$ and $\psi_1$ are defined by
\begin{equation}
F_1(s_1,s_2) = \E\big[e^{-s_1U_1-s_2U_2}\ind_{\{0< U_1 < U_2\}}\big], \; \; \; \; G_1(s_1) = \E\big[e^{-s_1U_1}\ind_{\{0=U_2<U_1\}}\big]
\label{FF}
\end{equation}
for $\mathbf{s} = (s_1, s_2) \in \overline{\mathbf{\Omega}}$; similarly, the Laplace transforms $F_2$ and $G_2$ of $\varphi_2$ and $\psi_2$ are defined by
\begin{equation}
F_2(s_1,s_2) = \E\big[e^{-s_1U_1-s_2U_2}\ind_{\{0< U_2 < U_1\}}\big], \; \; \; \; G_2(s_2) = \E\big[e^{-s_2U_2}\ind_{\{0=U_1<U_2\}}\big]
\label{FFb}
\end{equation}
for $\mathbf{s} = (s_1, s_2) \in \overline{\mathbf{\Omega}}$. Expression (\ref{dPhi}) for distribution $\mathrm{d}\Phi$ and the above definitions then enable to define the Laplace transform $F$ of the pair $(U_1,U_2)$ by
\begin{equation}
F(s_1,s_2) = 1 - \varrho + F_1(s_1,s_2) + G_1(s_1) + F_2(s_1,s_2) + G_2(s_2)
\label{Ldef}
\end{equation}
for $(s_1,s_2) \in \overline{\mathbf{\Omega}}$. Finally, let $b_1(s_1) = \mathbb{E}(e^{-s_1\mathcal{T}_1})$ (resp. $b_2(s_2) = \mathbb{E}(e^{-s_2\mathcal{T}_2})$) denote the Laplace transform of service time $\mathcal{T}_1$ (resp. $\mathcal{T}_2$) at queue $\sharp 1$ (resp. queue $\sharp 2$) 
for $\Re(s_1) \geq 0$ (resp. $\Re(s_2) \geq 0$); set in addition
\begin{equation}
\left\{
\begin{array}{ll}
K(s_1,s_2)=\lambda-\lambda_1 b_1(s_1)-\lambda_2 b_2(s_2), 
\\ \\
K_1(s_1,s_2) = s_1 - K(s_1,s_2), \; \; \; K_2(s_1,s_2) = s_2 - K(s_1,s_2)
\end{array} \right.
\label{Ker}
\end{equation}
and
\begin{equation}
\left\{
\begin{array}{ll}
J_1(s_1) = (1-\varrho)(\lambda - \lambda_1b_1(s_1)) - \psi_2(0),
\\ \\
J_2(s_2) = (1-\varrho)(\lambda - \lambda_2b_2(s_2)) - \psi_1(0).
\end{array} \right.
\label{J1J2}
\end{equation}
where the values $\psi_1(0)$, $\psi_2(0)$ at the origin of densities $\psi_1$, $\psi_2$ defined by \textbf{A.2}, �\ref{MA}, verify $
\psi_1(0) = \lim_{s_1\to \infty}s_1G_1(s_1)$ and $\psi_2(0) = \lim_{s_2\to \infty}s_2G_2(s_2)$, respectively. The following Proposition and Corollary are proved in [\textbf{Gui12}].
\begin{prop}
\textbf{a)} Transforms $F_1$, $G_1$ and $F_2$, $G_2$ together satisfy
\begin{equation}
K_1(s_1,s_2)H_1(s_1,s_2) + K_2(s_1,s_2)H_2(s_1,s_2) = (1-\varrho)K(s_1,s_2)
\label{Fonct}
\end{equation}
for $(s_1, s_2) \in \mathbf{\Omega}$, where $H_1 = F_1 + G_1$ and $H_2 = F_2 + G_2$.

\textbf{b)} Transforms $F_1$ and $G_2$ (resp. $F_2$, $G_1$) satisfy
\begin{equation}
\left\{ 
\begin{array}{lll}
K_1(s_1,s_2)F_1(s_1,s_2) + K_2(s_1,s_2) G_2(s_2) = J_2(s_2) + H(s_1,s_2),
\\
\\
K_2(s_1,s_2)F_2(s_1,s_2) + K_1(s_1,s_2) G_1(s_1) = J_1(s_1) - H(s_1,s_2)
\end{array} \right.
\label{Fonctcomp}
\end{equation}
for $(s_1, s_2) \in \mathbf{\Omega}$, with 
\begin{multline}
H(s_1,s_2) = \lambda_1 \mathbb{E} \left [ e^{-s_1U_1-s_2U_2}\ind_{\{0 \leq U_1 < U_2\}}e^{-s_1\mathcal{T}_1}\ind_{\{\mathcal{T}_1 > U_2-U_1\}} \right ] 
 \\ - \lambda_2\mathbb{E} \left [ e^{-s_1U_1-s_2U_2}\ind_{\{0 \leq U_2 < U_1\}}e^{-s_2\mathcal{T}_2}\ind_{\{\mathcal{T}_2 > U_1-U_2\}} \right ].
\label{H}
\end{multline}

\textbf{c)} Constants $\psi_1(0)$ and $\psi_2(0)$ satisfy relation $\psi_1(0) + \psi_2(0) = \lambda(1-\varrho)$.
\label{resol}
\end{prop}

\begin{corol} 
\label{coroltech} 
Let $H$ be defined by (\ref{H}). Transform $G_1$ satisfies
\begin{equation}
(s_1 - s_2)G_1(s_1) = J_1(s_1) - H(s_1,s_2)
\label{FonctG1}
\end{equation}
for $(s_1, s_2) \in \mathbf{\Omega}$ such that $K_2(s_1,s_2) = 0$. Similarly, transform $G_2$ satisfies
\begin{equation}
(s_2 - s_1)G_2(s_2) = J_2(s_2) + H(s_1,s_2)
\label{FonctG2}
\end{equation}
for $(s_1, s_2) \in \mathbf{\Omega}$ such that $K_1(s_1,s_2) = 0$.
\label{C.2}
\end{corol}

By Corollary~\ref{coroltech}, the determination of Laplace transforms $F_1$, $F_2$, $G_1$ and $G_2$ critically depends on the solutions to equations  $K_1(s_1,s_2)=0$ and $K_2(s_1,s_2)=0$. 

Analytic continuation properties are now stated as follows. Let $\overline{U}_j(t)$, $j \in \{1,2\}$, denote the workload in queue $\sharp j$ when the other queue has HoL priority; similarly, let $ \underline{U}_j(t)$ denote the workload in queue $\sharp j$ when this queue has HoL priority over the other. Assume that random variable $\overline{U}_j =\lim_{t \uparrow +\infty}\overline{U}_j(t)$ has an analytic Laplace transform $s \mapsto \E(e^{-s\overline{U}_j})$ in the domain $\{s \in \C \mid \; \Re(s )> \widetilde{s}_j\}$ for some real $\widetilde{s}_j<0$. By using stochastic domination arguments, we can show [\textbf{Gui12}] the following.
\begin{corol}
Laplace transform $F_1$ can be analytically extended to domain
$$
\widetilde{\mathbf{\Omega}}_1 = \{(s_1,s_2) \in \C^2 \mid \; \Re(s_2) > \max(\widetilde{s}_2,\widetilde{s}_2-\Re(s_1))\}
$$
and transform $G_2$ can be analytically extended to $\widetilde{\omega}_2 = \{s_2 \in \C \mid \; \Re(s_2)>\widetilde{s}_2\}$. 
\\
\indent
Similarly, transform $F_2$ can be analytically extended to
$$
\widetilde{\mathbf{\Omega}}_2 = \{(s_1,s_2) \in \C^2 \mid \; \Re(s_1) > \max(\widetilde{s}_1,\widetilde{s}_1-\Re(s_2)\}
$$
and $G_1$ can be analytically extended to $\widetilde{\omega}_1 = \{s_1\in \C \mid \; \Re(s_1)>\widetilde{s}_1\}$.
\label{extensions}
\end{corol} 
Actual values of $\widetilde{s}_1$ and $\widetilde{s}_2$ are specified below in Lemma \ref{HoLpoles} in the case of exponentially distributed service times.


\section{Exponential service times}
\label{EXPI}


In the following, we will specify the discussion to the case when the service time at queue $\sharp j \in \{1,2\}$ is exponentially distributed with parameter $\mu_j > 0$, associated with Laplace transforms $b_j(s) = \mu_j/(s +\mu_j)$ for $\Re(s) \geq 0$. Expression~(\ref{Ker}) for $K(s_1,s_2)$ presently reads
\begin{equation}
K(s_1,s_2) = \frac{\lambda_1s_1}{s_1+\mu_1} + \frac{\lambda_2s_2}{s_2+\mu_2}.
\label{Kexp}
\end{equation}
This first section is dedicated to key preparatory facts for the derivation of Laplace transforms $F_1$, $F_2$ and $G_1$, $G_2$.

\subsection{Zeros of kernels}
\label{ZEK}

As mentioned above, it is essential to compute the zeros of kernels $K_1$ and $K_2$ introduced in (\ref{Ker}). Analytic together with geometric characterisations of such  zeros can then be formulated as follows.
\begin{lemma}
\textbf{a)} For given $s_1 \in \mathbb{C}$, equation $K_2(s_1,s_2) = 0$ has two solutions
\begin{equation}
s_2 = \xi_2^+(s_1), \; \; s_2 = \xi_2^-(s_1)
\label{xi2+-0}
\end{equation}
such that $\xi_2^+(0) = 0$ and $\xi_2^-(0) = \lambda_2 - \mu_2$; complex function $\xi_2^-$ (resp. $\xi_2^+$) has an analytic (resp. meromorphic) extension to the cut plane $\mathbb{C} \setminus [\zeta_1^-,\zeta_1^+]$, where
\begin{equation}
\zeta_1^- = -\mu_1 \frac{(\sqrt{\mu_2}+\sqrt{\lambda_2})^2}{\lambda_1+(\sqrt{\mu_2}+\sqrt{\lambda_2})^2}, \quad \zeta_1^+ = -\mu_1 \frac{(\sqrt{\mu_2}-\sqrt{\lambda_2})^2}{\lambda_1+(\sqrt{\mu_2}-\sqrt{\lambda_2})^2}.
\label{zeta1+-}
\end{equation}
\indent
In a similar manner, equation $K_1(s_1,s_2) = 0$ for given $s_2 \in \mathbb{C}$ has two solutions
\begin{equation}
s_1 = \xi_1^+(s_2), \; \; s_1 = \xi_1^-(s_2)
\label{xi1+-0}
\end{equation}
such that $\xi_1^+(0) = 0$ and $\xi_1^-(0) = \lambda_1 - \mu_1$; function $\xi_1^-$ (resp. $\xi_1^+$) has an analytic (resp. meromorphic) extension to the cut plane $\mathbb{C} \setminus [\zeta_2^-,\zeta_2^+]$, where $\zeta_2^\pm$ are defined from  (\ref{zeta1+-}) by permuting indexes 1 and 2.
\\
\indent
\textbf{b)} For $\mu_1 \neq \mu_2$, the non-zero roots of equation $K(s,s) = s$ are that of quadratic
\begin{equation}
P(s) = s^2+(\mu_1+\mu_2-\lambda)s+\mu_1\mu_2(1-\varrho)
\label{PolP}
\end{equation}
with $\lambda = \lambda_1 + \lambda_2$; these roots are real negative, say,  $\sigma_0^- < \sigma_0^+ < 0$ with
$$
\sigma_0^\pm = \frac{-(\mu_1+\mu_2-\lambda) \pm \sqrt{(\mu_1-\mu_2-\lambda_1+\lambda_2)^2+4\lambda_1\lambda_2}}{2}.
$$
For $\mu_1 = \mu_2$, the only non-zero root of equation $K(s,s) = s$ is $\sigma_0 = -\mu(1-\varrho)$.
\label{Roots0}
\end{lemma}
\begin{proof}
\textbf{a)} Given $s_1 \in \mathbb{C}$, definition (\ref{Ker}) and specific expression (\ref{Kexp}) of $K(s_1,s_2)$ imply that equation $K_2(s_1,s_2)=0$ is quadratic in variable $s_2$ and expands as $(s_1+\mu_1)s_2^2+\left(\mu_1\mu_2-\lambda_2\mu_1+(\mu_2-\lambda_1-\lambda_2)s_1\right)s_2-\lambda_1\mu_2s_1 = 0$, 
which provides two solutions explicitly given by  
\begin{equation}
\xi^{\pm}_2(s_1)  = \frac{-\left((\mu_2-\lambda_2)\mu_1+(\mu_2-\lambda_1-\lambda_2)s_1\right)\pm\sqrt{D_1(s_1)}}{2(s_1+\mu_1)},
\label{xi2+-}
\end{equation}
with discriminant $D_1(s_1) = \left(\mu_1\mu_2-\lambda_2\mu_1+(\mu_2-\lambda_1-\lambda_2)s_1\right)^2+4\lambda_1\mu_2s_1(\mu_1+s_1)$ and where $\sqrt{u}$ denotes the analytic determination of the square root on the cut plane $\mathbb{C} \setminus \mathbb{R}^-$ such that $\sqrt{u} > 0$ for $u > 0$. Simple computations show that $D_1(s_1)$ vanishes at $s_1 = \zeta_1^-$ and $s_1 = \zeta_1^+$ given by (\ref{zeta1+-}), which define two ramification points for functions $\xi_2^\pm$. Furthermore, as verified in Appendix \ref{A30}, $\xi_2^-$ (resp. $\xi_2^+$) can be extended as an analytic (resp. meromorphic) function on the cut plane $\C \setminus[\zeta_1^-,\zeta_1^+]$, the point $s_1 = -\mu_1$ being a pole for $\xi_2^+$ only. 
\\
\indent
\emph{Mutatis mutandis}, the analysis of equation $K_1(s_1,s_2)=0$ in variable $s_1$ similarly defines functions $\xi_1^\pm$ given by
\begin{equation}
\xi^{\pm}_1(s_2)  = \frac{-\left((\mu_1-\lambda_1)\mu_2+(\mu_1-\lambda_1-\lambda_2)s_2\right)\pm\sqrt{D_2(s_2)}}{2(s_2+\mu_2)},
\label{xi1+-}
\end{equation}
with $D_2(s_2) = \left(\mu_1\mu_2-\lambda_1\mu_2+(\mu_1-\lambda_1-\lambda_2)s_2\right)^2+4\lambda_2\mu_1s_2(\mu_2+s_2)$.
\\
\indent
\textbf{b)} When $\mu_1\neq \mu_2$, equation $s= K(s,s)$ is readily seen to be equivalent to quadratic equation $P(s) = 0$, with quadratic polynomial $P$ defined by (\ref{PolP}). The discriminant of $P(s)$ is positive since it equals $(\mu_1 - \mu_2 - \lambda_1 + \lambda_2)^2 + 4\lambda_1\lambda_2$; $P$ has therefore two real roots with negative sum and positive product (after stability condition $\varrho < 1$), implying that these roots are negative 
The case $\mu_1=\mu_2$ is immediate.
\end{proof}

\subsection{Geometric aspects}
\label{GA}

While equation $K_2(s_1,s_2) = 0$ (resp. $K_1(s_1,s_2) = 0$) has degree 2 in variable $s_2$ (resp. $s_1)$, it has only degree 1 in variable $s_1$ (resp. $s_2$) and has the unique solution
\begin{equation}
s_1 = T_2(s_2) \quad \mathrm{(resp.} \quad s_2 = T_1(s_1) \mathrm{)}
\label{T1T2}
\end{equation}
where
$$
T_2(s_2) = - \frac{\mu_1s_2(s_2+\mu_2-\lambda_2)}{s_2^2 - (\lambda - \mu_2)s_2 - \lambda_1\mu_2}, \; \; 
T_1(s_1) = - \frac{\mu_2s_1(s_1+\mu_1-\lambda_1)}{s_1^2 - (\lambda - \mu_1)s_1 - \lambda_2\mu_1}
$$
define rational functions with order 2; by construction, identities $T_1 \circ \xi_1^\pm = \mathrm{Id}$ and $T_2 \circ \xi_2^\pm = \mathrm{Id}$ hold.
\begin{prop}
Let $\widehat{\mathbb{C}} = \mathbb{C} \cup \{\infty\}$ denote the Riemann sphere (i.e., the compactified complex plane). 

Then the algebraic curve $\mathcal{R}_1$ (resp. algebraic curve $\mathcal{R}_2$) in $\widehat{\mathbb{C}} \times \widehat{\mathbb{C}}$ defined by equation $K_1(s_1,s_2) = 0$ (resp. $K_2(s_1,s_2) = 0$) is a cubic with genus 0.
\label{Genus0}
\end{prop}
\begin{proof}
As $K_1(s_1,s_2) = 0 \Leftrightarrow s_2 = T_1(s_1)$ by (\ref{T1T2}), curve $\mathcal{R}_1$ is a rational cubic since rational function $T_1$ has order 2. As a rational curve, it has therefore genus 0 [\textbf{Fis01}, �9.3], that is, it is homeomorphic to the Riemann sphere itself. Algebraic representation (\ref{xi1+-0}) and rational representation (\ref{T1T2}) are, in particular, topologically equivalent. The same conclusions hold for curve $\mathcal{R}_2$.
\end{proof}
\begin{figure}[b]
\scalebox{1}{\includegraphics[width=12cm,trim=120 310 30 90 cm,clip]{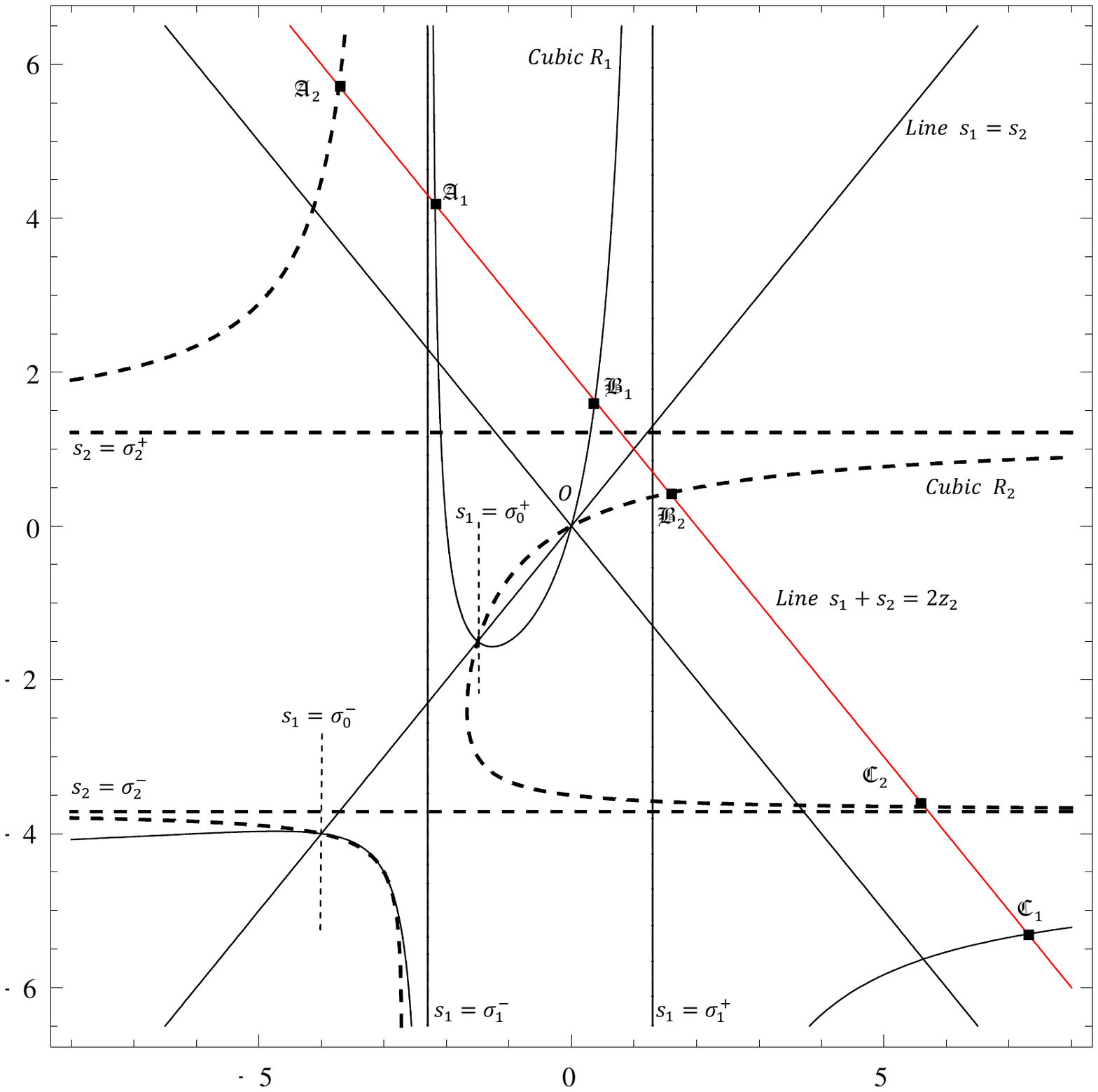}}
\caption{\textit{Cubic $\mathcal{R}_1$, full line (resp. cubic $\mathcal{R}_2$, dashed line) with line $s_1 + s_2 = 2z$ (red line) in the real $(O,s_1,s_2)$ plane.}}
\label{Fig4}
\end{figure}
Proposition \ref{Genus0} stresses the fact that the rationality of cubics $\mathcal{R}_1$ and $\mathcal{R}_2$ is quite specific, since a cubic is  generically non rational with genus 1 [\textbf{Fis01},�9.7]. 

Cubics $\mathcal{R}_1$ and $\mathcal{R}_2$ are illustrated in the real $(O,s_1,s_2)$ plane for the specific values $\lambda_1 = 1$, $\lambda_2 = 1$, $\mu_1 = 3$, $\mu_2 = 4.5$ (Fig.\ref{Fig4}). Generally, rational functions $T_1$ and $T_2$ have simple poles $\sigma_1^- < 0 < \sigma_1^+$ and $\sigma_2^- < 0 < \sigma_2^+$, respectively, with
\begin{equation}
\sigma_1^{\pm} = \frac{\lambda-\mu_1 \pm \sqrt{(\lambda-\mu_1)^2 + 4\lambda_2\mu_1}}{2}, \;
\sigma_2^{\pm} = \frac{\lambda-\mu_2 \pm \sqrt{(\lambda-\mu_2)^2 + 4\lambda_1\mu_2}}{2}
\label{poles1s2}
\end{equation}
so that $\mathcal{R}_1$ (resp. $\mathcal{R}_2$) has vertical (resp. horizontal) asymptotes at $s_1 = \sigma_1^\pm$ (resp. at $s_2 = \sigma_2^\pm$). Cubic $\mathcal{R}_1$ also has horizontal tangents at stationary points $s_1$ verifying $T_1'(s_1) = 0$; differentiating expression (\ref{T1T2}) of $T_1(s_1)$, the latter equation has solutions $s_1 = a_1^\pm $ with
\begin{equation}
a_1^\pm = -\mu_1 \pm \sqrt{\lambda_1\mu_1}.
\label{statpoints1}
\end{equation}
Similarly, $\mathcal{R}_2$ has vertical tangents at stationary points $s_2$ verifying $T_2'(s_2) = 0$, hence $s_2 = a_2^\pm$ with
\begin{equation}
a_2^\pm = -\mu_2 \pm \sqrt{\lambda_2\mu_2}.
\label{statpoints2}
\end{equation}
\indent
By Lemma \ref{Roots0}.b and when $\mu_1 \neq \mu_2$, the non-zero roots of equation $K(s,s) = s$ are that of polynomial $P(s)$ defined in (\ref{PolP}); apart from the origin $O = (0,0)$, cubics $\mathcal{R}_1$ and $\mathcal{R}_2$ therefore meet at points $(\sigma_0^-,\sigma_0^-)$ and $(\sigma_0^+,\sigma_0^+)$; when $\mu_1 = \mu_2$, they meet at $O$ and at point $(\sigma_0,\sigma_0)$ only, where $\sigma_0 = \lambda - \mu$.
\\
\indent
Finally, ramification points  $\zeta_1^\pm$ of functions $\xi_2^\pm$ defined by (\ref{zeta1+-}) are equivalently the values of $s_1$ for which equations $T_2(s_2) = s_1$ and $T_2'(s_2) = 0$ have a common solution $s_2$; from definition (\ref{statpoints2}), this implies that $s_2 = a_2^\pm$ and consequently $\zeta_1^\pm = T_2(a_2^\pm)$. As $a_2^\pm$ is a local minimum of $T_2$ on an interval including $\sigma_0^+$, we deduce that $\zeta_1^+ = T_2(a_2^+) \leq T_2(\sigma_0^+) = \sigma_0^+$, hence inequalities
\begin{equation}
\zeta_1^- < \zeta_1^+ \leq \sigma_0^+.
\label{zeta1-ineq}
\end{equation}
We similarly have $\zeta_2^\pm = T_1(a_1^\pm)$ and $\zeta_2^- < \zeta_2^+ \leq \sigma_0^+$ for the ramification points $\zeta_2^\pm$ of functions $\xi_1^\pm$.

\subsection{Analytic continuation}
\label{MAD}

\begin{center}
\begin{figure}[t]
\scalebox{1}{\includegraphics[width=12cm,trim=120 60 30 80 cm,clip]{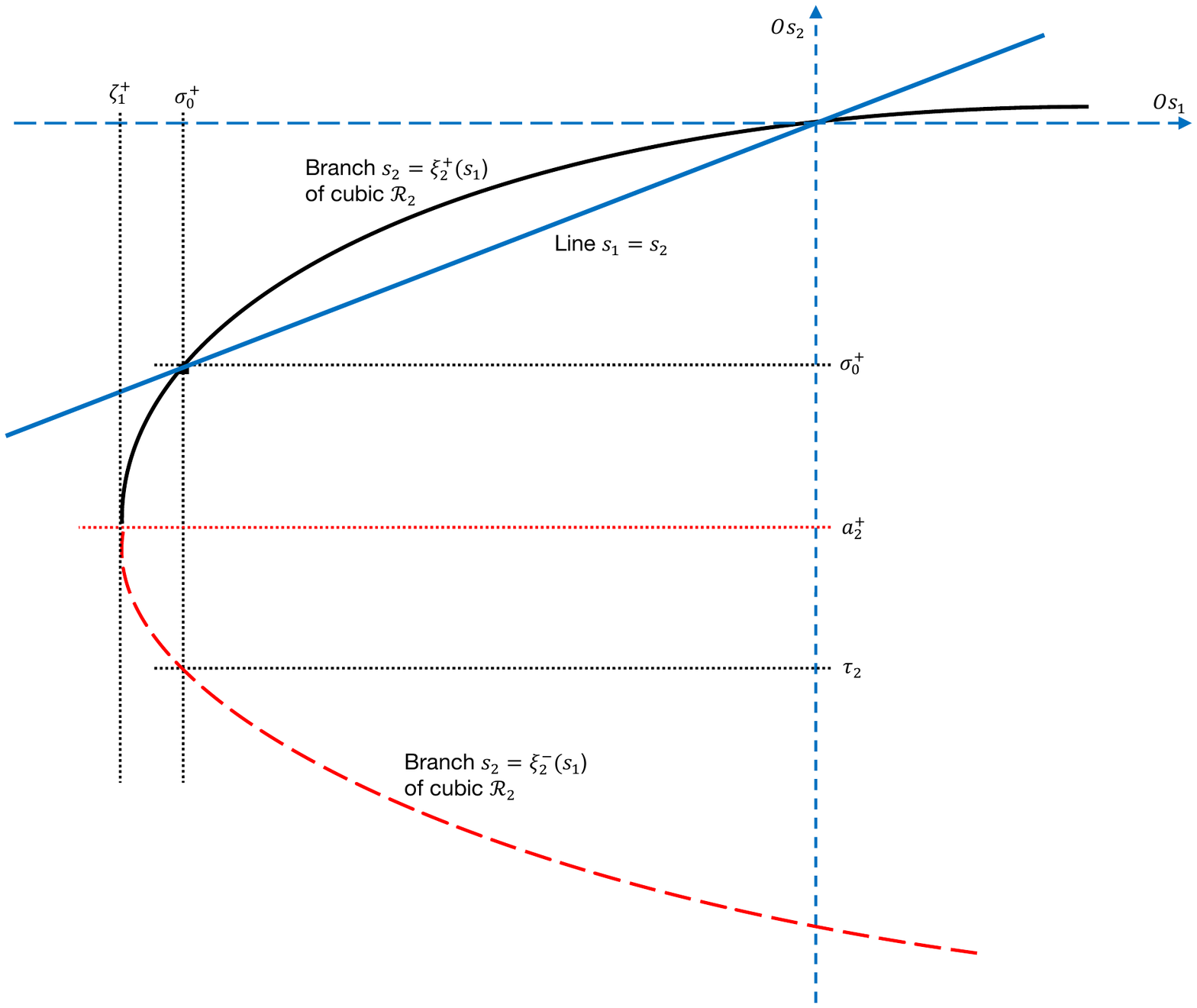}}
\caption{\textit{Branches $s_2=\xi_2^+(s_1)$ (black line) and $s_2 = \xi_2^-(s_1)$ (red dashed line) of cubic $\mathcal{R}_2$ and their intersection with line $s_1 = s_2$ in the real $(O,s_1,s_2)$ plane - assuming condition $(I^+)$ to hold.}}
\label{Fig9}
\end{figure}
\end{center}

We now specify the extended analyticity domains of tranforms $F_1$, $G_2$ (resp. $F_2$, $G_1$) in the case of exponential service time distributions; following Corollary \ref{extensions}, this amounts to explicit $\widetilde{s}_1$ and $\widetilde{s}_2$. It is known [\textbf{Gui04}, �3.3] that when queue $\sharp$2 has HoL priority over queue $\sharp$1, the Laplace transform of the workload $\overline{U}_1$ of queue $\sharp 1$ 
is given by
\begin{equation}
\E\big[e^{-s \overline{U}_1}\big] = \frac{(1-\rho)s_1\xi_2^+(s_1)}{\lambda_1(1-b_1(s_1))(s_1-\xi_2^+(s_1))}, \; \; \Re(s_1) \geq 0,
\label{HoL21}
\end{equation}
where $\xi_2^+$ is defined by \eqref{xi2+-}; its analyticity domain is now specified as follows. 
\begin{lemma}
Transform $s_1 \mapsto \E(e^{-s \overline{U}_1})$ is analytic in $\{s \in \C \mid \; \Re(s)>\widetilde{s}_1\}$ where 
$$
\widetilde{s}_1 = \left\{\begin{array}{ll} \sigma_0^+ & \mbox{\; if \; \; } (2\mu_2-\mu_1)\sqrt{\varrho_2}+\mu_1 \leq \mu_2 + \lambda_1 + \lambda_2 \; \; \; \; \; \; \mathrm{(}I^+\mathrm{)},
\\ \\
 \zeta_1^+ & \mbox{\; if \; \; } (2\mu_2-\mu_1)\sqrt{\varrho_2}+\mu_1 >  \mu_2+\lambda_1+\lambda_2 \; \; \; \; \; \; \mathrm{(}I^-\mathrm{)}.
\end{array}  \right.
$$
Similarly, transform $s_2 \mapsto \E(e^{-s_2 \overline{U}_2})$ is analytic in $\{s \in \C : \Re(s)>\widetilde{s}_2\}$ where 
$$
\widetilde{s}_2 = \left\{\begin{array}{ll} \sigma_0^+ & \mbox{\; if \; \; } (2\mu_1-\mu_2)\sqrt{\varrho_1}+\mu_2 \leq \mu_1+\lambda_1+\lambda_2 \; \; \; \; \; \; \mathrm{(}II^+\mathrm{)},
\\ \\
 \zeta_2^+ & \mbox{\; if \; \; } (2\mu_1-\mu_2)\sqrt{\varrho_1}+\mu_2 >  \mu_1+\lambda_1+\lambda_2 \; \; \; \; \; \; \mathrm{(}II^-\mathrm{)}.
\end{array}  \right.
$$
\label{HoLpoles}
\end{lemma}
\begin{proof}
By Lemma \ref{Roots0}.a, expression (\ref{HoL21}) defines a meromorphic transform in the cut plane $\C\setminus[\zeta_1^-,\zeta_1^+]$; its possible poles are the solutions to $s_1=\xi_2^+(s_1)$, that is, $s_1 =\sigma_0^\pm$ as 
defined in Lemma \ref{Roots0}.b (recall by inequalities (\ref{zeta1-ineq}) that $\zeta_1^+ \leq \sigma_0^+$).
\\
\indent
To localise such poles, consider the "upper" and "lower" branches $s_2 = \xi_2^+(s_1)$ and $s_2 = \xi_2^-(s_1)$ of cubic $\mathcal{R}_2$ for $s_1 > \zeta_1^+$. In the real $(O,s_1,s_2)$ plane, these branches and the vertical axis $Os_2$ delineate a convex domain and the straight line $s_1 = s_2$ interesects that domain at either $s_1=s_2=0$ or $s_1 = s_2 = \sigma_0^+$ (see Fig.\ref{Fig9} for illustration). The latter intersection point then belongs to the upper branch $s_2 = \xi_2^+(s_1)$ if and only if $\sigma_0^+ \geq a_2^+$, in which case $\sigma_0^+$ is a pole for expression (\ref{HoL21}). Conversely, condition $\sigma_0^+ < a_2^+$ ensures that $\sigma_0^+$ is not a pole for (\ref{HoL21}) and that its smallest singularity is consequently $\zeta_1^+$.
\\
\indent
By \eqref{PolP}, quadratic $P(s)$ has roots $\sigma_0^- < \sigma_0^+ < 0$.  Condition $\sigma_0^+ \geq a_2^+$ is then equivalent to $P (a_2^+) \leq 0$, which reduces to  $(2\mu_2-\mu_1)\sqrt{\varrho_2}+\mu_1-\mu_2-\lambda_1-\lambda_2 \leq 0$. We can then conclude that threshold $\widetilde{s}_1$ equals either $\sigma_0^+$ or $\zeta_1^+$ according to condition $(I^+)$ or $(I^-)$, as claimed. A similar proof holds for threshold $\widetilde{s}_2$. 
\end{proof}

\subsection{Auxiliary functions}

We conclude this section by showing that the determination of bivariate transforms $F_1$ and $F_2$ to that of some univariate functions $M_1$ and $M_2$. This fact  actually follows from the following proposition [\textbf{Gui12}, Proposition 3.3].
\begin{prop}
In the case of exponentially distributed service times, function $H$ defined in \eqref{H} explicitly reads
\begin{equation}
H(s_1,s_2) =\frac{\lambda_1\mu_1}{\mu_1+s_1} M_1\left(\frac{s_1+s_2}{2}\right) - \frac{\lambda_2\mu_2}{\mu_2+s_2} M_2\left(\frac{s_1+s_2}{2}\right)
\label{Hbis}
\end{equation}
where
$$
\left\{
\begin{array}{ll}
M_1(z) = G_2(2z+\mu_1)+F_1(-\mu_1,2z+\mu_1),
\\ \\
M_2(z) = G_1(2z+\mu_2)+F_2(2z+\mu_2,-\mu_2)
\end{array} \right.
$$
are analytically defined for $\Re(z) > \max(\widetilde{s}_1,\widetilde{s}_2)/2$, with thresholds $\widetilde{s}_1$ and $\widetilde{s}_2$ given in Lemma \ref{HoLpoles}.
\label{Hexp}
\end{prop}
Once function $H$ has been expressed as in Proposition \ref{Hexp}, simply solving each equation (\ref{Fonctcomp}) for $F_1$ and $F_2$ gives 
\begin{align}
\left\{ 
\begin{array}{ll}
F_1(s_1,s_2) = \displaystyle \frac{J_2(s_2) - K_2(s_1,s_2)G_2(s_2)}{K_1(s_1,s_2)} + \frac{H(s_1,s_2)}{K_1(s_1,s_2)}, 
\\ \\
F_2(s_1,s_2)  = \displaystyle \frac{J_1(s_1) - K_1(s_1,s_2)G_1(s_1)}{K_2(s_1,s_2)} - \frac{H(s_1,s_2)}{K_2(s_1,s_2)}
\end{array} \right.
\label{f1f2bis}
\end{align}
for $(s_1,s_2) \in \mathbf{\Omega}$, respectively. As univariate transforms $G_1$ and $G_2$ will also be shown to depend on auxiliary functions $M_1$ and $M_2$ only, our remaining task is therefore to derive the latter functions. 


\section{Solving functionals equations}
\label{EXPII}


Once the latter algebraic and analytic results have been stated, the objective of this section is to show that functions $M_1$ and $M_2$ verify a two-dimensional functional equation that is subsequently solved.

\subsection{Properties of cubics}

Before stating our main result in �\ref{FEG1G2}, complementary properties of kernels $K_1$ and $K_2$ must be formulated. In fact, formula (\ref{Hbis}) for $H(s_1,s_2)$ motivates the introduction of the variable change $(s_1,s_2) \mapsto (w,z)$ where
\begin{equation}
2 w = s_1 - s_2, \; \; \; 2 z = s_1 + s_2.
\label{varch}
\end{equation}
In the rest of this paper, we define
$$
R_1(w,z) = w^3 + \sum_{k=1}^3 R_{1k}w^{3-k}, \; \; \; R_2(w,z) = w^3 + \sum_{k=1}^3 R_{2k}w^{3-k}
$$ 
as the cubic polynomials with coefficients
\begin{equation}
\left\{ 
\begin{array}{ll} 
R_{11}(z) = - (\lambda-\mu_1+\mu_2-z), \\
R_{12}(z) = \lambda_1\mu_2-\lambda_2\mu_1-\mu_1\mu_2-2\mu_2z-z^2, \\
R_{13}(z) = - zP(z),
\end{array} \right. 
\label{R1}
\end{equation}
and
\begin{equation}
\left\{ 
\begin{array}{ll} 
R_{21}(z) = \lambda+\mu_1-\mu_2-z, \\
R_{22}(z) = \lambda_2\mu_1-\lambda_1\mu_2-\mu_1\mu_2-2\mu_1z-z^2, \\
R_{23}(z) = zP(z),
\end{array} \right. 
\label{R2}
\end{equation}
respectively, where $P(z)$ is given by (\ref{PolP}).
\begin{lemma} (see Proof in Appendix \ref{A3}) 
\\ \indent
\textbf{a)} Kernels $K_1(s_1,s_2) = K_1(z+w,z-w)$ and $K_2(s_1,s_2) = K_2(z+w,z-w)$ defined in (\ref{Ker}) read 
\begin{equation}
\left\{
\begin{array}{ll}
K_1(z+w,z-w) = \displaystyle - \frac{R_1(w,z)}{(w + z + \mu_1)(-w+z+\mu_2)}, 
\\
\\
K_2(z+w,z-w) = \displaystyle + \frac{R_2(w,z)}{(w + z + \mu_1)(-w+z+\mu_2)}
\end{array} \right.
\label{abexp}
\end{equation}
with cubics $R_1(w,z)$ and $R_2(w,z)$ given by (\ref{R1}) and (\ref{R2}), respectively. They further verify identity
\begin{equation}
R_1(w,z) + R_2(w,z) = -2w(w+z+\mu_1)(-w+z+\mu_2).
\label{R1R2}
\end{equation}
\indent
\textbf{b)} For given $z > 0$, polynomial $R_1(\centerdot,z)$ has 3 real roots 
$\alpha_1(z)$, $\beta_1(z)$, $\gamma_1(z)$ such that $\alpha_1(z) < -z < \beta_1(z) < z < \gamma_1(z)$. 
\\
\indent
Similarly, polynomial $R_2(\centerdot,z)$ has 3 real roots $\alpha_2(z)$, $\beta_2(z)$, $\gamma_2(z)$ such that $\alpha_2(z) < -z < \beta_2(z) < z < \gamma_2(z)$.
\label{Roots}
\end{lemma}
Fix some $z > 0$ (if not mentioned, the dependence of roots $\alpha_1$, $\beta_1$, $\gamma_1$ and $\alpha_2$, $\beta_2$, $\gamma_2$ on variable $z$ is implicit). By Lemma \ref{Roots}.b, the intersection points $\mathfrak{A}_1$, $\mathfrak{B}_1$, $\mathfrak{C}_1$ of cubic $\mathcal{R}_1$ with line $s_1 + s_2 = 2z$ respectively are related to the roots $w = \alpha_1$, $w = \beta_1$, $w = \gamma_1$ of equation $R_1(w,z) = 0$ (see Fig.\ref{Fig4} for illustration). By variable change (\ref{varch}), the coordinates $(a_1,A_1)$, $(b_1,B_1)$, $(c_1,C_1)$ of $\mathfrak{A}_1, \mathfrak{B}_1, \mathfrak{C}_1$ in the $(O,s_1,s_2)$ frame are thus  given by
\begin{equation}
\left\{
\begin{array}{ll}
a_1 = z + \alpha_1, \; \; \; b_1 = z + \beta_1, \; \; \; c_1 = z + \gamma_1,
\\ \\
A_1 = z - \alpha_1, \; \; B_1 = z - \beta_1, \; \; C_1 = z - \gamma_1,
\end{array} \right.
\label{coord1}
\end{equation}
respectively. Symmetrically, the intersection points $\mathfrak{A}_2, \mathfrak{B}_2, \mathfrak{C}_2$ of cubic $\mathcal{R}_2$ with the same line $s_1 + s_2 = 2z$ (see Fig.\ref{Fig4} for illustration) have coordinates $(A_2,a_2)$, $(B_2,b_2)$, $(C_2,c_2)$ in the $(O,s_1,s_2)$ frame given by
\begin{equation}
\left\{
\begin{array}{ll}
A_2 = z + \alpha_2, \; \; B_2 = z + \beta_2, \; \; C_2 = z + \gamma_2,
\\ \\
a_2 = z - \alpha_2, \; \; \; b_2 = z - \beta_2, \; \; \; c_2 = z - \gamma_2
\end{array} \right.
\label{coord2}
\end{equation}
(note the exchange of lower and upper case between (\ref{coord1}) and (\ref{coord2}) is due to the exchange of variables $s_1$ and $s_2$ when passing from cubic $\mathcal{R}_1$ to cubic $\mathcal{R}_2$).
\\
\indent
\begin{figure}[t]
\scalebox{1}{\includegraphics[width=13cm, trim = 100 480 0 90 cm,clip]{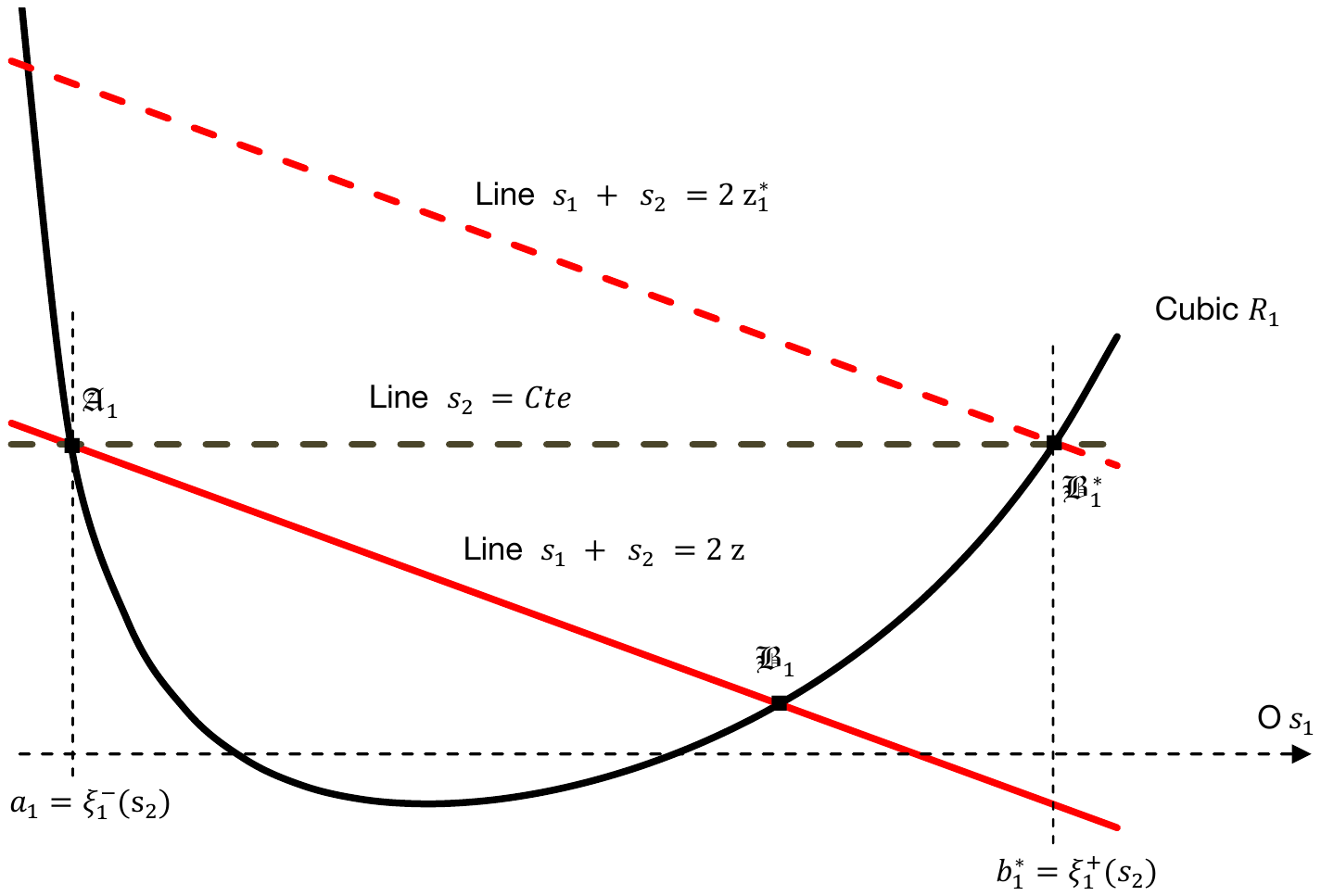}}
\caption{\textit{Line $L: s_1 + s_2 = 2z$ crosses cubic $\mathcal{R}_1$ at points $\mathfrak{A}_1, \mathfrak{B}_1$. Parallel line $L_1^*: s_1 + s_2 = 2z_1^*$, $z_1^* > z$, crosses $\mathcal{R}_1$ at point $\mathfrak{B}_1^*$ with same ordinate 
than $\mathfrak{A}_1$.}}
\label{Fig5}
\end{figure}
A simple geometric operation on cubics $\mathcal{R}_1$ and $\mathcal{R}_2$  will prove essential for the determination of auxiliary functions $M_1$ and $M_2$. For given $z > 0$, let $\mathfrak{A}_1$ and $\mathfrak{B}_1^*$ denote the two finite intersections of cubic $\mathcal{R}_1$ with the line $D$ passing through $\mathfrak{A}_1$ and parallel to axis $Os_1$ (see Fig.\ref{Fig5}); as the third intersection point of $\mathcal{R}_1$ with $D$ is at infinity in the $Os_1$ direction, the point $\mathfrak{B}_1^*$ is uniquely determined once  $\mathfrak{A}_1$ is given. Let $L$ denote the line with equation $s_1 + s_2 = 2z$; the line $L_1^*$ parallel to $L$ and passing through $\mathfrak{B}_1^*$ has then equation $s_1 + s_2 = 2z_1^*$ for some $z_1^* > 0$. By the previous definition of $\mathfrak{B}_1^*$, we have
\begin{equation}
\forall \; z > 0, \; \; A_1 = B_1^*
\label{intersect}
\end{equation}
where ordinates $A_1 = A_1(z)$ and $B_1^* = B_1(z_1^*)$ are defined by  (\ref{coord1}). 
\begin{lemma} (see Proof in Appendix \ref{A6}) 
\\
\indent
For any $z > 0$, let $a_1 = a_1(z)$ (resp. $b_1^* = b_1(z_1^*)$) denote the abcissa of point $\mathfrak{A}_1$ (resp. of point $\mathfrak{B}_1^*$) of cubic $\mathcal{R}_1$ with $\sigma_1^- < a_1 < b_1^* < \sigma_1^+$. We have 
\begin{equation}
b_1^* = -\mu_1 \frac{a_1+\mu_1-\lambda_1}{a_1+\mu_1}.
\label{condition}
\end{equation}
Defining function $h_1$ by $h_1(z) = z_1^*$, we further have
\begin{equation}
2h_1(z) = 2z - (a_1+\mu_1) + \frac{\lambda_1\mu_1}{a_1 + \mu_1}
\label{z2z2*}
\end{equation}
and $h_1(z) > z$ for all $z > 0$.
\label{iteration0}
\end{lemma}
\noindent 
Lemma \ref{iteration0} actually asserts the invariance property $T_1 \circ \iota_1 = T_1$ of rational function $T_1$ with respect to transformation $\iota_1:a_1 \mapsto b_1^*$. As specified by (\ref{condition}), transformation $\iota_1$ is rational and is readily verified to be an involution on $\mathcal{R}_1$, that is, $\iota_1 \circ \iota_1 = \iota_1$. The rationality of $\iota_1$ is consistent with the general L�roth theorem [\textbf{Bel09}, �8.8.2, p.275], ensuring that any involution on a rational curve $\mathcal{R}_1$ must be rational.

The notation $h_1$ here refers to the fact that $h_1(z)$ measures some kind of "height" of point on curve $\mathcal{R}_1$; analytic properties of $h_1$ are formulated below in �\ref{ACM} below. Similar observations hold for function $h_2$ related to curve $\mathcal{R}_2$.
\\ \\
\indent
Let then $s_2 \geq 0$ and fix the ordinate $A_1 = z - \alpha_1(z) = s_2$ of the intersection point $\mathfrak{A}_1$ with smallest abcissa $a_1$ (that is, $a_1 < b_1 < c_1$). In relation to algebraic functions $\xi_1^\pm$ defined in Lemma \ref{Roots0}, we then have $a_1 = \xi_1^-(s_2)$; by equations (\ref{coord1}), the coordinates of point $\mathfrak{A}_1$ in the $(O,w,z)$ frame are
\begin{equation}
w_1 = \alpha_1(z), \; \; z = \frac{\xi_1^-(s_2)+s_2}{2}
\label{coord1a}
\end{equation}
with $s_2 = z - \alpha_1(z)$. Fixing the \textit{same ordinate} $B_1^* = z_1^* - \beta_1(z_1^*) = s_2$, the abcissa $b_1^*$ of its image $\mathfrak{B}_1^*$ then equals $b_1^* = \xi_1^+(s_2)$; correspondingly, the coordinates of $\mathfrak{B}_1^*$ in the $(O,w,z)$ frame read
\begin{equation}
w_1^* = \beta_1(z^*), \; \; z_1^* = \frac{\xi_1^+(s_2)+s_2}{2}
\label{coord1b*}
\end{equation}
with $z_1^* = h_1(z)$. 
\\
\indent
Similarly, the image of intersection point $\mathfrak{C}_2 = (C_2,c_2)$ on cubic $\mathcal{R}_2$ is defined by $\mathfrak{B}_2^{*} = (B_2^{*},b_2^{*})$ with identical abcissa  $C_2 = B_2^{*}$ in the $(O,s_1,s_2)$ frame (with notation (\ref{coord2})). Such a point $\mathfrak{B}_2^{*}$ is on the line with equation $s_1 + s_2 = 2z_2^{*}$; this enables us to define function $h_2$ by $h_2(z) = z_2^{*}$ with
\begin{equation}
2h_2(z) = 2z - (c_2+\mu_2) + \frac{\lambda_2\mu_2}{c_2 + \mu_2}.
\label{z2z2**}
\end{equation}
Let $s_1 \geq 0$ and fix the abcissa $C_2 = z + \gamma_2(z) = s_1$ of the intersection point $\mathfrak{C}_2$ with smallest ordinate (that is, $c_2 < b_2 < a_2$); the coordinates of $\mathfrak{C}_2$ in the $(O,w,z)$ frame are 
\begin{equation}
w_2 = \gamma_2(z), \; \; z = \frac{\xi_2^-(s_1)+s_1}{2}
\label{coord2c}
\end{equation}
with $s_1 = z + \gamma_2(z)$, while the coordinates of its image $\mathfrak{B}_2^{*}$ are
\begin{equation}
w_2^{*} = \beta_2(z_2^{*}), \; \; z_2^{*} = \frac{\xi_2^+(s_1)+s_1}{2}
\label{coord2b*}
\end{equation}
with $z_2^{*} = h_2(z)$.


\subsection{Functional equations for $M_1$ and $M_2$}
\label{FEG1G2}


We now specify the functional equations verified by functions $M_1$ and $M_2$ and complete their resolution. 

Consider any root $\epsilon_j(z) \in \{\alpha_j(z), \beta_j(z), \gamma_j(z)\}$ of polynomial $R_j(\cdot,z)$, $j \in \{1,2\}$, defined in Lemma \ref{Roots}.a for real $z > 0$. We let
\begin{equation}
q_1(z;\epsilon_j(z)) = \frac{\lambda_1\mu_1}{\mu_1 + z + \epsilon_j(z)}, \; \; q_2(z;\epsilon_j(z)) = \frac{\lambda_2\mu_2}{\mu_2 + z - \epsilon_j(z)}
\label{q_epsilon}
\end{equation}
and simply write $q_1(z;\epsilon_j(z)) = q_1(\epsilon_j)$ and $q_2(z;\epsilon_j(z)) = q_2(\epsilon_j)$ without mentioning the current argument $z$ of $\epsilon_j$. Given $\beta_1 = \beta_1(z)$ and $\beta_2 = \beta_2(z)$, we also set
$$
\beta_1^* = \beta_1(z_1^*), \; \; \; \beta_2^{*} = \beta_2(z_2^{*})
$$
where $z_1^* = h_1(z)$ and $z_2^{*} = h_2(z)$ are defined by (\ref{z2z2*}) and (\ref{z2z2**}), respectively; we similarly write $q_1(\beta_1^*) = q_1(z_1^*,\beta_1(z_1^*))$ and $q_2(\beta_2^{*}) = q_2(z_2^{*},\beta_2(z_2^{*}))$.
\begin{prop} 
Consider the $2 \times 1$ column vector $\mathbf{M}(z) = \left (M_1(z) \; \; M_2(z)\right)^T$. Then $\mathbf{M}$ verifies the two-dimensional functional equation
\begin{equation}
\mathbf{M}(z) = Q_1(z) \cdot \mathbf{M} \circ h_1(z) +Q_2(z) \cdot \mathbf{M} \circ h_2(z) + \mathbf{L}(z)
\label{FonctG1G2-equ}
\end{equation}
for all $z > 0$, where $2 \times 2$ matrices $Q_1 = k_1 \Pi_1$ and $Q_2 = k_2 \Pi_2$ are defined by factors
$$
k_1(z) = \frac{1}{D(z)}\frac{s_2-\xi_1^-(s_2)}{s_2-\xi_1^+(s_2)}, \; \; k_2(z) = \frac{1}{D(z)}\frac{s_1-\xi_2^-(s_1)}{s_1-\xi_2^+(s_1)}
$$
with $s_2 = z - \alpha_1(z)$, $s_1 = z + \gamma_2(z)$ and 
\begin{equation}
D(z) = 4\lambda_1\mu_1\lambda_2\mu_2  \frac{(\mu_1+\mu_2+2z)\alpha_1\gamma_2(\alpha_1-\gamma_2)}{R_1(\gamma_2,z)R_2(\alpha_1,z)},
\label{detD}
\end{equation} 
by matrices
$$
\Pi_1 = \begin{pmatrix}
-q_2(\gamma_2) q_1(\beta_1^*) & q_2(\gamma_2) q_2(\alpha_1) \\ \\
-q_1(\gamma_2) q_1(\beta_1^*) & q_1(\gamma_2) q_2(\alpha_1)
\end{pmatrix}, \; \; \; \; 
\Pi_2 = \begin{pmatrix}
q_2(\alpha_1) q_1(\gamma_2) & -q_2(\alpha_1) q_2(\beta_2^{**}) \\ \\
q_1(\alpha_1) q_1(\gamma_2) & -q_1(\alpha_1) q_2(\beta_2^{**})
\end{pmatrix}
$$
and where the $2 \times 1$ column vector $\mathbf{L}(z) = \left( L_1(z) \; \; L_2(z) \right)^T$ is given by
$$
L_1 = \frac{1}{D}\left[ q_2(\alpha_1) \left ( \frac{\xi_2^-(s_1)-\xi_2^+(s_1)}{s_1-\xi_2^+(s_1)}\right )J_1(s_1) - q_2(\gamma_2)\left ( \frac{\xi_1^+(s_2)-\xi_1^-(s_2)}{s_2-\xi_1^+(s_2)}\right ) J_2(s_2) \right],
$$
$$
L_2 = \frac{1}{D}\left[ q_1(\alpha_1) \left ( \frac{\xi_2^-(s_1)-\xi_2^+(s_1)}{s_1-\xi_2^+(s_1)}\right )J_1(s_1) - q_1(\gamma_2)\left ( \frac{\xi_1^+(s_2)-\xi_1^-(s_2)}{s_2-\xi_1^+(s_2)}\right ) J_2(s_2) \right].
$$
\label{FonctG1G2}
\end{prop}
\begin{proof} 
Observe that $s_2+\xi_1^+(s_2)$ and $s_2+\xi_1^-(s_2)$ are positive for large enough real $s_2$; equation (\ref{FonctG2}) of Corollary~\ref{coroltech} therefore applies to $s_1 = \xi_1^+(s_2)$ and $s_1 = \xi_1^-(s_2)$, respectively. Using (\ref{coord1a})-(\ref{coord1b*}), we thus obtain 
$$
\left\{
\begin{array}{ll}
(s_2-\xi_1^+(s_2))G_2(s_2) = J_2(s_2)+ \displaystyle  \frac{\lambda_1\mu_1}{\mu_1+\xi_1^+(s_2)} M_1(h_1(z)) - 
\frac{\lambda_2\mu_2}{\mu_2+s_2} M_2(h_1(z)),
\nonumber \\ \\
(s_2-\xi_1^-(s_2))G_2(s_2) = J_2(s_2)+ \displaystyle  \frac{\lambda_1\mu_1}{\mu_1+\xi_1^-(s_2)} M_1(z) - \frac{\lambda_2\mu_2}{\mu_2+s_2} M_2(z)
\end{array} \right.
$$
for large enough real $s_2$, with $s_2 = z - \alpha_1$ and $\xi_1^-(s_2) = z + \alpha_1$. 
Equating then the common value of $G_2(s_2)$ from the above equations, we have
\begin{multline}
\label{eqM22}
q_1(\alpha_1) M_1(z) - q_2(\alpha_1) M_2(z) = \displaystyle \frac{\xi_1^+(s_2)-\xi_1^+(s_2)}{s_2-\xi_1^+(s_2)}J_2(s_2) \; + \\
\frac{\lambda_1\mu_1(s_2-\xi_1^-(s_2))}{(\mu_1+\xi_1^+(s_2))(s_2-\xi_1^+(s_2))} M_1(h_1(z)) - \displaystyle \frac{\lambda_2\mu_2(s_2-\xi_1^-(s_2))}{(\mu_2+s_2)(s_2-\xi_1^+(s_2))} M_2(h_1(z))
\end{multline}
for large enough real $s_2$ and with $q_1(\alpha_1)$ and $q_2(\alpha_1)$ defined in (\ref{q_epsilon}) for $\epsilon_1 = \alpha_1$. 
\\
\indent
Similarly, we note that $s_1+\xi_2^+(s_1) \geq 0$ and $s_1+\xi_2^-(s_1)\geq 0$ for large enough real $s_1$; equation~\eqref{FonctG1} therefore applies to $s_2 = \xi_2^+(s_1)$ and $s_2 = \xi_2^-(s_1)$, respectively. Using (\ref{coord2c})-(\ref{coord2b*}), we thus obtain
$$
\left\{
\begin{array}{ll}
(s_1-\xi_2^+(s_1))G_1(s_1) = \displaystyle J_1(s_1) -  \frac{\lambda_1\mu_1}{\mu_1+s_1} M_1\left(h_2(z)\right) + \frac{\lambda_2\mu_2}{\mu_2+\xi_2^+(s_1)} M_2\left(h_2(z)\right),
\nonumber \\ \\
(s_1-\xi_2^-(s_1))G_1(s_1) = \displaystyle J_1(s_1) -  \frac{\lambda_1\mu_1}{\mu_1+s_1} M_1(z) + \frac{\lambda_2\mu_2}{\mu_2+\xi_2^-(s_1)} M_2(z)
\end{array} \right.
$$
for large enough real $s_1$, with $s_1 = z + \gamma_2$ and $\xi_2^-(s_1) = z - \gamma_2$. 
Equating the common value of $G_1(s_1)$ from the above equations, we thus obtain
\begin{multline}
\label{eqM11}
q_1(\gamma_2) M_1(z) - q_2(\gamma_2) M_2(z)  = \frac{\xi_2^-(s_1)-\xi_2^+(s_1)}{s_1-\xi_2^+(s_1)} J_1(s_1) \; + \\ 
\frac{\lambda_1\mu_1(s_1-\xi_2^-(s_1))}{(\mu_1+s_1)(s_1-\xi_2^+(s_1))} M_1(h_2(z)) - \frac{\lambda_2\mu_2(s_1-\xi_2^-(s_1))}{(\mu_2+\xi_2^+(s_1))(s_1-\xi_2^+(s_1))} M_2(h_2(z))
\end{multline}
for 
$z > 0$ and with $q_1(\gamma_2)$ and $q_2(\gamma_2)$ defined in (\ref{q_epsilon}) for  $\epsilon_2 = \gamma_2$.
Now, equations (\ref{eqM22})-(\ref{eqM11}) define a linear system
\begin{equation}
V\mathbf{M}(z) = \mathbf{N}_0 + V_1\mathbf{M}(h_1(z)) + V_2 \mathbf{M}(h_2(z))
\label{LM}
\end{equation}
with $2 \times 2$ matrix 
$$
V = \begin{pmatrix}
q_1(\alpha_1) & -q_2(\alpha_1) \\ \\
q_1(\gamma_2) & -q_2(\gamma_2)
\end{pmatrix},
$$
some diagonal matrices $V_1$, $V_2$ and some $2 \times 1$ vector $\mathbf{N}_0$; that system can then be solved for vector $\mathbf{M}(z)$ in terms of $\mathbf{M}(h_1(z))$ and $\mathbf{M}(h_2(z))$, provided that matrix $V$ above has non-zero determinant $D = \mathrm{det}V$. In fact, applying definition (\ref{q_epsilon}) for coefficients $q_1(\gamma_2)$, $q_2(\gamma_2)$ and $q_1(\alpha_1)$, $q_2(\alpha_1)$, we calculate
\begin{align}
D & = q_1(\gamma_2)q_2(\alpha_1) - q_2(\gamma_2)q_1(\alpha_1)
\nonumber \\
& = \frac{\lambda_1\mu_1}{\mu_1+\gamma_2+z}\frac{\lambda_2\mu_2}{\mu_2-\alpha_1+z} - \frac{\lambda_2\mu_2}{\mu_2-\gamma_2+z} \frac{\lambda_1\mu_1}{\mu_1+\alpha_1+z}
\nonumber \\
& = \lambda_1\mu_1\lambda_2\mu_2 \frac{(\mu_1+\mu_2+2z)(\alpha_1-\gamma_2)}{(\mu_1+\gamma_2+z)(\mu_2-\gamma_2+z)(\mu_1+\alpha_1+z)(\mu_2-\alpha_1+z)};
\nonumber
\end{align}
use relation (\ref{R1R2}) for $R_1 + R_2$ to write 
\begin{equation}
(\mu_1+\alpha_1+z)(\mu_2-\alpha_1+z) = - \frac{R_2(\alpha_1,z)}{2\alpha_1}
\label{R2alpha1}
\end{equation}
since $R_1(\alpha_1,z) = 0$; we similarly write
\begin{equation}
(\mu_1+\gamma_2+z)(\mu_2-\gamma_2+z) = - \frac{R_1(\gamma_2,z)}{2\gamma_2};
\label{R1gamma2}
\end{equation}
determinant $D = \mathrm{det}V$ then reduces to expression (\ref{detD}) and is consequently non-zero for $z > 0$ in view of Lemma \ref{Roots}.b. Solving then system (\ref{LM}) for $\mathbf{M}(z)$ in terms of $\mathbf{M}(h_1(z))$ and $\mathbf{M}(h_2(z))$ readily provides functional relation (\ref{FonctG1G2-equ}).
\end{proof}
As derived in the proof of Proposition \ref{FonctG1G2}, transforms $G_1$ and $G_2$ are now expressed in terms of auxiliary functions $M_1$ and $M_2$ either by
\begin{equation}
\left\{
\begin{array}{ll}
G_1(s_1) = \displaystyle \frac{1}{s_1-\xi_2^+(s_1)} \left [ J_1(s_1) -  \frac{\lambda_1\mu_1M_1 (h_2(z))}{\mu_1+s_1}  + \frac{\lambda_2\mu_2M_2( h_2(z))}{\mu_2+\xi_2^+(s_1)}  \right ],
\\ \\
G_2(s_2) = \displaystyle \frac{1}{s_2-\xi_1^+(s_2)} \left [ J_2(s_2)+  \frac{\lambda_1\mu_1M_1(h_1(z))}{\mu_1+\xi_1^+(s_2)}  - \frac{\lambda_2\mu_2M_2 (h_1(z))}{\mu_2+s_2}  \right]
\end{array} \right.
\label{G1G2-M1M2+}
\end{equation}
or by
\begin{equation}
\left\{
\begin{array}{ll}
G_1(s_1) = \displaystyle \frac{1}{s_1-\xi_2^-(s_1)} \left [ J_1(s_1) -  \frac{\lambda_1\mu_1M_1(z)}{\mu_1+s_1}  + \frac{\lambda_2\mu_2M_2(z)}{\mu_2+\xi_2^-(s_1)}  \right ],
\\ \\
G_2(s_2) = \displaystyle  \frac{1}{s_2-\xi_1^-(s_2)} \left [ J_2(s_2)+  \frac{\lambda_1\mu_1M_1(z)}{\mu_1+\xi_1^-(s_2)}  - \frac{\lambda_2\mu_2M_2(z)}{\mu_2+s_2}  \right ]
\end{array} \right.
\label{G1G2-M1M2-}
\end{equation}
for large enough real $s_1 = z + \gamma_2(z)$ and $s_2 = z - \alpha_1(z)$ so that $z > 0$. We are now left to solve functional equation (\ref{FonctG1G2-equ}) for $M_1$ and $M_2$.
\\
\indent
Let then $< h_1, h_2 >$ denote the semi-group (equipped with the function composition operation) generated by $h_1$ and $h_2$, that is, the set of all compositions $h = h_{i_1} \circ h_{i_2} \circ ... \circ h_{i_k}$ for any $k \in \mathbb{N}$ and $(i_1,...,i_k) \in \{1,2\}^k$ (by convention, we set $h = \mathrm{Id}$ for $k = 0$). The elements of semi-group $< h_1, h_2 >$ can be represented as the nodes of the infinite binary tree with root the identity mapping $\mathrm{Id}$, and where each element $h$ has children $h \circ h_1$ and $h \circ h_2$. We now assert the central result of this section.
\begin{theorem}
With the above notation for semi-group $< h_1, h_2>$, the column vector $\mathbf{M} = \left (M_1 \; \; M_2 \right)^T$ is given by the series expansion
\begin{equation}
\mathbf{M}(z) = \sum_{k=0}^{+\infty} \; \sum_{(i_1,...,i_k) \in \{1,2\}^k} \prod_{\ell = 0}^{k-1} Q_{i_{k-\ell}}(h_{i_{k-\ell+1},...,i_k}) \cdot \mathbf{L}  \circ h_{i_1, ..., i_k}(z)
\label{M1M2series}
\end{equation}
for all $z > 0$, with $h_{i,j, ...,\ell} = h_i \circ h_j \circ ... \circ h_\ell$ and where
$$
\prod_{\ell = 0}^{k-1} Q_{i_{k-\ell}}(h_{i_{k-\ell+1},...,i_k}) = Q_{i_k}Q_{i_{k-1}}(h_{i_k})...Q_{i_1}(h_{i_2, ..., i_k})
$$
is a product matrix, with matrices $Q_1$ and $Q_2$ introduced in Proposition \ref{FonctG1G2} (by convention, that product reduces to the unit matrix $\mathrm{Id}$ for $k = 0$, and we set $Q_{i_{k-\ell}}(h_{i_{k-\ell+1},...,i_k}) = \mathrm{Id}$ for $k \geq 1$ and $\ell = 0$).
\label{FonctM1M2}
\end{theorem}
\begin{proof}
For given $z > 0$, let $\mathbf{M}_k(z)$ denote the generic term at order $k \geq 0$ of series (\ref{M1M2series}). Apply then recursively functional equation (\ref{FonctG1G2-equ}) to order $K \geq 1$ to obtain
$$
\mathbf{M}(z) = \sum_{0 \leq k \leq K} \mathbf{M}_k(z) + \mathbf{E}^{(K)}(z)
$$
with remainder 
$$
\mathbf{E}^{(K)}(z) = \sum_{(i_1,...,i_{K+1}) \in \{1,2\}^{K+1}} \prod_{\ell = 0}^{K} Q_{i_{K+1-\ell}}(h_{i_{K-\ell+2},...,i_{K+1}}) \cdot \mathbf{M} \circ h_{i_1,...,i_K,i_{K+1}}(z).
$$
\indent
We now show that $\mathbf{E}^{(K)}(z) \rightarrow 0$ as $K \uparrow +\infty$. As shown in [\textbf{Gui12}], Theorem 5.1, the sequence of iterated $h_1 \circ ... \circ h_1(z)$, $K$ times, (resp. $h_2 \circ ... \circ h_2(z)$, $K$ times) of function $h_1$ (resp. function $h_2$) tends to $+\infty$ when $K \uparrow +\infty$. As a consequence, any iterated $h_{i_1,...,i_K,i_{K+1}}(z)$ tends to $+\infty$ when $K \uparrow +\infty$. On the other hand, following definition (\ref{Hbis}),  functions $M_1$ and $M_2$ are bounded in the neighborhood of infinity since $F_1$, $F_2$ and $G_1$, $G_2$ all vanish at infinity as transforms of regular densities; the sequence $\mathbf{M} \circ h_{i_1,...,i_K,i_{K+1}}(z)$, $(i_1,...,i_K, i_{K+1}) \in \{1,2\}^{K+1}$, $K \geq 0$, is consequently bounded.
\\
\indent
By arguments similar to that of [\textbf{Gui12}], Theorem 5.1, abcissa $a_1 = a_1(Z)$ (resp. ordinate $A_1 = A_1(Z)$) when calculated at argument $Z$ tending to $+\infty$ tends to $\sigma_1^-$ (resp. $+\infty$); abcissa $b_1^* = b_1(h_1(Z))$ (resp. ordinate $b_2^{*} = b_2(h_2(Z))$) tends to $\sigma_1^+$ (resp. to $\sigma_2^+$); and finally, abcissa $C_2 = C_2(Z)$ (resp. ordinate $c_2 = c_2(Z)$) tends to $+\infty$ (resp. to $\sigma_2^-$). From the above observations, identities (\ref{coord1})-(\ref{coord2}) and definitions (\ref{q_epsilon}) of functions $q_1$ and $q_2$ together imply
\begin{align}
& q_1(\alpha_1) = \frac{\lambda_1\mu_1}{\mu_1 + a_1} \rightarrow \frac{\lambda_1\mu_1}{\mu_1 + \sigma_1^-}, \; \; \; q_2(\alpha_1) = \frac{\lambda_2\mu_2}{\mu_2 + A_1} \rightarrow 0,
\nonumber \\
& q_1(\beta_1^*) = \frac{\lambda_1\mu_1}{\mu_1 + b_1^*} \rightarrow \frac{\lambda_1\mu_1}{\mu_1 + \sigma_1^+}, \; \; \; q_2(\beta_2^{*}) = \frac{\lambda_2\mu_2}{\mu_2 + b_2^{*}} \rightarrow \frac{\lambda_2\mu_2}{\mu_2 + \sigma_2^+},
\nonumber \\
& q_1(\gamma_2) = \frac{\lambda_1\mu_1}{\mu_1 + C_2} \rightarrow 0, \; \; \; \; \; \; \; \; \; \; \; \; \; \; q_2(\gamma_2) = \frac{\lambda_2\mu_2}{\mu_2 + c_2} \rightarrow \frac{\lambda_2\mu_2}{\mu_2 + \sigma_2^-}
\nonumber
\end{align}
when $Z \uparrow +\infty$. From the expressions of matrices $\Pi_1 = \Pi_1(Z)$ and $\Pi_2 = \Pi_2(Z)$ given in Proposition \ref{FonctG1G2}, the above results enables us to deduce that
$$
\Pi_1(Z) \rightarrow \begin{pmatrix}
- \displaystyle \frac{\lambda_1\mu_1}{\mu_1 + \sigma_1^+} \times  \frac{\lambda_2\mu_2}{\mu_2 + \sigma_2^-} & 0 \\
0 & 0  
\end{pmatrix}, \; \; \; \Pi_2(Z) \rightarrow 
\begin{pmatrix}
0 & 0 \\
0 & \displaystyle - \frac{\lambda_1\mu_1}{\mu_1 + \sigma_1^-} \times  \frac{\lambda_2\mu_2}{\mu_2 + \sigma_2^+}
\end{pmatrix}
$$
as $Z \uparrow +\infty$. On the other hand, the definitions of factors $k_1(Z)$ and $k_2(Z)$ given in Proposition \ref{FonctG1G2} give in turn
$$
D(Z)k_1(Z) = \frac{s_2-a_1}{s_2-b_1^*} \rightarrow 1, \; \; \; D(Z)k_2(Z) = \frac{s_1 - c_2}{s_1 - b_2^{*}} \rightarrow 1
$$
as $s_2 = Z - \alpha_1(Z) \rightarrow +\infty$, $s_1 = Z + \gamma_2(Z) \rightarrow +\infty$ when $Z \uparrow +\infty$ and $a_1 \rightarrow \sigma_1^-$, $b_1 \rightarrow \sigma_1^+$, $c_2 \rightarrow \sigma_2^-$, $b_2^{*} \rightarrow \sigma_2^+$. Besides, asymptotics $a_1 = Z + \alpha_1 \sim \sigma_1^-$ and $c_2 = Z - \gamma_2 \sim \sigma_2^-$ give $\alpha_1 \sim -Z$ and $\gamma_2 \sim Z$ for large positive $Z$; we then deduce from identities (\ref{R2alpha1})-(\ref{R1gamma2}) that $R_2(\alpha_1,Z) \sim -(-2Z)(\mu_1+\sigma_1^-)2Z$ and $R_1(\gamma_2,Z) \sim -(2Z)2Z(\mu_2+\sigma_2^-)$. Using the above estimates, definition (\ref{detD}) of $D(Z)$ then gives
$$
D(Z) \sim 4\lambda_1\mu_1\lambda_2\mu_2 \frac{2Z \times (-Z)Z(-2Z)}{(-4(\mu_2+\sigma_2^-)Z^2)(4(\mu_1+\sigma_1^-)Z^2)} = - \frac{\lambda_1\lambda_2\mu_1\mu_2}{(\mu_1+\sigma_1^-)(\mu_2+\sigma_2^-)}.
$$
The previous estimates therefore show that matrices $Q_1(Z) = k_1(Z)\Pi_1(Z)$ and  $Q_2(Z) = k_2(Z)\Pi_2(Z)$ tend to 
$$ 
\begin{pmatrix}
r_1 & 0 \\
\displaystyle 0 & \displaystyle 0
\end{pmatrix}, \; \; \; \; \; \; 
\begin{pmatrix}
0 & 0 \\
0 & r_2
\end{pmatrix}
$$
respectively, where $r_j =(\mu_j + \sigma_j^-)/(\mu_j + \sigma_j^+)$. The non-zero element $r_j$ is clearly positive and $< 1$ since $0 < \mu_j + \sigma_j^- < \mu_j + \sigma_j^+$ for each $j \in \{1,2\}$. Using explicit expressions (\ref{poles1s2}) of $\sigma_1^-$ and $\sigma_1^+$, we further note that
$$
r_1 = \frac{\mu_1+\sigma_1^-}{\mu_1+\sigma_1^+} = \frac{4\varrho_1}{(\varrho_1+m\varrho_2+1 + \sqrt{(\varrho_1+m\varrho_2-1)^2+4m\varrho_2})^2}
$$
is a decreasing function of the ratio $m = \mu_2/\mu_1$, and equals $\varrho_1$ for $m = 0$; we thus deduce that $r_1 \leq \varrho_1$ and similarly $r_2 \leq \varrho_2$. Coming back to the definition of remainder $\mathbf{E}^{(K)}(z)$ above, the above arguments therefore imply that
$$
\mathbf{E}^{(K)}(z) = O \left ( \sum_{n_1+n_2 = K+1} r_1^{n_1}r_2^{n_2} \right ) = O(r_1 + r_2)^K = O(\varrho^K)
$$
for large $K$, where $\varrho = \varrho_1 + \varrho_2 < 1$. Remainder $\mathbf{E}^{(K)}(z)$ therefore tends to 0 for increasing $K$; as $\mathbf{M}(z)$ is finite for any $z > 0$ by the existence of the stationary distribution, we conclude that series expansion (\ref{M1M2series}) holds for such values of $z$. 
\end{proof}
Following (\ref{M1M2series}), solution $\mathbf{M}$ linearly depends on vector $\mathbf{L}$ and is therefore a linear combination of functions $J_1$ and $J_2$ introduced in (\ref{J1J2}). The latter still depend on unknown constants $\psi_1(0)$ and $\psi_2(0)$ which can be determined as follows. First write vector $\mathbf{L}(z) = \mathbf{L}^{(0)}(z)+\psi_1(0) \mathbf{L}^{(1)}(z)+\psi_2(0)\mathbf{L}^{(2)}(z)$ as a linear combination of $\psi_1(0)$ and $\psi_2(0)$ with
$$
\left\{
\begin{array}{ll}
\mathbf{L}^{(0)}(z) =  \displaystyle \frac{-(1-\varrho)}{D(z)}\left[   \frac{\xi_2^+(s_1)-\xi_2^-(s_1)}{s_1-\xi_2^+(s_1)}(\lambda-\lambda_1b_1(s_1)) \mathbf{e}_2(z) \; + \right. \\ \\
\left. \; \; \; \; \; \; \; \; \; \; \; \; \; \; \; \; \; \; \; \; \; \; \; \; \; \; \; \; \; \; \; \; \displaystyle \frac{\xi_1^+(s_2)-\xi_1^-(s_2)}{ s_2-\xi_1^+(s_2)}(\lambda-\lambda_2b_2(s_2))\mathbf{e}_1(z) \right],
\\ \\
\mathbf{L}^{(1)}(z) = \displaystyle \frac{1}{D(z)}  \frac{\xi_1^+(s_2)-\xi_1^-(s_2)}{ s_2-\xi_1^+(s_2)}\mathbf{e}_1(z), \; \; \mathbf{L}^{(2)}(z) = \displaystyle \frac{1}{D(z)}  \frac{\xi_2^+(s_1)-\xi_2^-(s_1)}{s_1-\xi_2^+(s_1)}\mathbf{e}_2(z)
\end{array} \right.
$$
with the $2 \times 1$ column vectors $\mathbf{e}_1(z) = (q_2(\gamma_2) \; \; q_1(\gamma_2))^T$, $\mathbf{e}_2(z) = (q_2(\alpha_1) \; \; q_1(\alpha_1))^T$ and where $s_1=z+\gamma_2(z)$, $s_2=z-\alpha_1(z)$. For $i \in \{0,1,2\}$, let now  $\mathcal{L}^{(i)}$ denote the $2 \times 1$ vector satisfying the functional equation 
$$
\mathcal{L}^{(i)}(z) = Q_1(z) \cdot \mathcal{L}^{(i)} \circ h_1(z) +Q_2(z) \cdot  \mathcal{L}^{(i)} \circ h_2(z) + \mathbf{L}^{(i)}(z)
$$
for $z>0$, whose solution is given by Theorem \ref{FonctM1M2} as
\begin{equation}
\mathcal{L}^{(i)}(z) = \sum_{k=0}^{+\infty} \; \sum_{(i_1,...,i_k) \in \{1,2\}^k} \prod_{\ell = 0}^{k-1} Q_{i_{k-\ell}}(h_{i_{k-\ell+1},...,i_k}) \cdot \mathbf{L}^{(i)}  \circ h_{i_1, ..., i_k}(z)
\label{Li}
\end{equation}
so that $\mathbf{M} = \mathcal{L}^{(0)} +\psi_1(0) \mathcal{L}^{(1)} + \psi_2(0)\mathcal{L}^{(2)}$.
\begin{prop}
For each $i \in \{0,1,2\}$, denote by $\mathcal{L}^{(i)}_j$,  $j \in \{1,2\}$, the components of vector $\mathcal{L}^{(i)}(z)$ defined by expansion (\ref{Li}).
\\
\indent
\textbf{a)} Constants $\psi_1(0)$ and $\psi_2(0)$ are then given by
\begin{multline}
\psi_1(0) = \frac{\lambda_1(1-\varrho)+\lambda_1\mathcal{L}_1^{(0)}(0) -\lambda_2   \mathcal{L}_2^{(0)}(0)  +\lambda(1-\varrho) [ \lambda_1  \mathcal{L}_1^{(2)}(0) -\lambda_2\mathcal{L}_2^{(2)}(0) ]}{1-\lambda_1  \mathcal{L}_1^{(1)}(0)+ \lambda_1 \mathcal{L}_1^{(2)}(0) +\lambda_2 \mathcal{L}_2^{(1)}(0)-\lambda_2 \mathcal{L}_2^{(2)}(0)}
\nonumber 
\end{multline} 
and
\begin{multline}
\psi_2(0) = \frac{\lambda_2(1-\varrho)-\lambda_1\mathcal{L}_1^{(0)}(0) + \lambda_2   \mathcal{L}_2^{(0)}(0) - \lambda(1-\varrho) [\lambda_1  \mathcal{L}_1^{(1)}(0) - \lambda_2\mathcal{L}_2^{(1)}(0) ]}{1-\lambda_1  \mathcal{L}_1^{(1)}(0)+ \lambda_1 \mathcal{L}_1^{(2)}(0) +\lambda_2 \mathcal{L}_2^{(1)}(0)-\lambda_2 \mathcal{L}_2^{(2)}(0)}.
\nonumber
\end{multline}
\indent
\textbf{b)} The empty queue probabilities are given by
\begin{equation}
\mathbb{P}(U_1=0) = 1 - \varrho + G_2(0), \; \; \; \mathbb{P}(U_2=0) = 1 - \varrho + G_1(0)
\label{Empty12}
\end{equation}
with
$$
G_2(0) = \lim_{s_2 \downarrow 0} \displaystyle \frac{1}{s_2-\xi_1^+(s_2)} \left [ J_2(s_2)+  \frac{\lambda_1\mu_1M_1(h_1(z))}{\mu_1+\xi_1^+(s_2)}  - \frac{\lambda_2\mu_2M_2(h_1(z))}{\mu_2+s_2}  \right]
$$
where $h_1(z) = (s_2 + \xi_1^+(s_2))/2$, and
$$
G_1(0) = \lim_{s_1 \downarrow 0} \displaystyle \frac{1}{s_1-\xi_2^+(s_1)} \left [ J_1(s_1) -  \frac{\lambda_1\mu_1M_1(h_2(z))}{\mu_1+s_1}  + \frac{\lambda_2\mu_2M_2(h_2(z))}{\mu_2+\xi_2^+(s_1)} \right ]
$$
where $h_2(z) = (s_1 + \xi_2^+(s_1))/2$, respectively.
\label{psi12}
\end{prop}
\begin{proof}
\textbf{a)} Following (\ref{Hbis}), we have $H(0,0) = \lambda_1M_1(0) - \lambda_2M_2(0)$; besides, (\ref{J1J2}) gives $J_2(0) = \lambda_1(1-\varrho) - \psi_1(0)$. Appying equation (\ref{FonctG2}) for $s_1 = s_2 = 0$ and invoking the finiteness of $G_2(0)$ consequently implies that $J_2(0) + H(0,0) = 0$. Reduce then the latter equation to $\lambda_1 M_1(0) - \lambda_2 M_2(0) = \psi_1(0) - \lambda_1(1-\varrho)$ and combine it with identity $\psi_1(0) + \psi_2(0) = \lambda(1-\varrho)$ of Proposition \ref{resol}.c; solving for both $\psi_1(0)$ and $\psi_2(0)$ provides the announced formulas.
\\
\indent
\textbf{b)} Writing $\mathbb{P}(U_1 = 0) = \mathbb{P}(U_1 = U_2 = 0) + \mathbb{P}(U_1 = 0 < U_2)$ together with $\mathbb{P}(U_1 = U_2 = 0) = 1 - \varrho$, identity (\ref{Empty12}) follows by definition (\ref{FFb}) of $G_2$. Now, to calculate $G_2(0)$, apply relation (\ref{G1G2-M1M2+}) for $G_2(s_2)$ with $s_2 = 0$; as $\xi_1^+(0) = 0$ by (\ref{xi1+-}) and since $G_2(0)$ is finite, $G_2(0)$ is necessarily equal to the limit of the quotient expressed above. \emph{Mutatis mutandis}, the same derivation pattern holds for $\mathbb{P}(U_2 = 0)$ and $G_1(0)$. 
\end{proof}


\begin{figure}[b]
\scalebox{1}{\includegraphics[width=20cm, trim = 70 200 0 90 cm,clip]{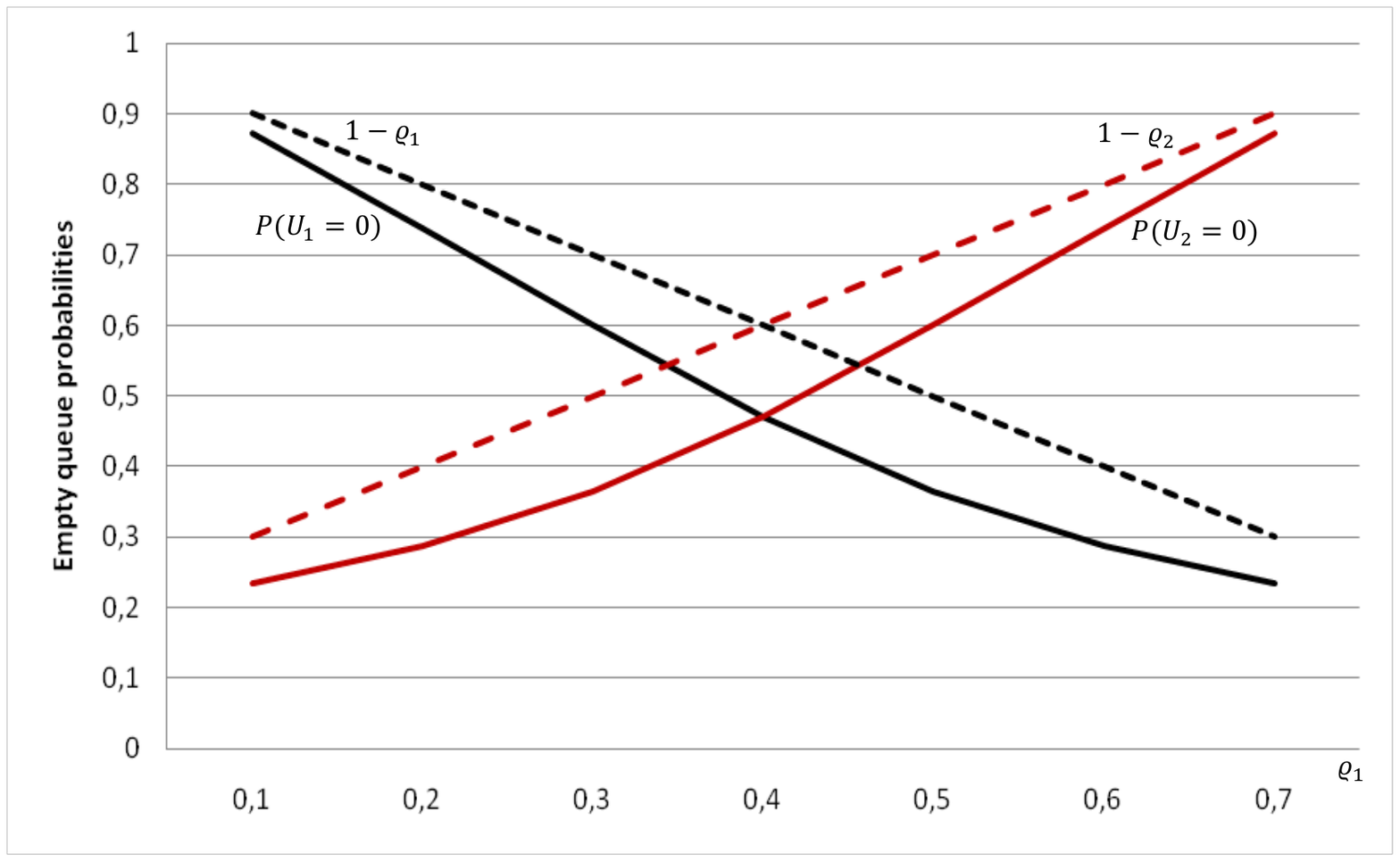}}
\caption{\textit{Empty queue probabilities $\mathbb{P}(U_1 = 0)$ and $\mathbb{P}(U_2 = 0)$ as functions of load $\varrho_1$, assuming $\mu_1 = \mu_2 = 1$ and $\varrho_1 + \varrho_2 = 0.8$.}}
\label{Fig5ter}
\end{figure}
In Fig.\ref{Fig5ter}, we depict the variations of empty queue probabilities $\mathbb{P}(U_1 = 0)$ and $\mathbb{P}(U_2 = 0)$ as a function of $\varrho_1$, assuming the total load $\varrho$ is fixed. Implementing formulae of Proposition \ref{psi12} was easily performed under Mathematica software tool by using tree structures, as numerous iterations are necessary for computing infinite sums and products. We note that for small load $\varrho_1$, probability $\mathbb{P}(U_1 = 0)$ is close enough to probability $1 - \varrho_1$ that would be obtained if a fixed HoL priority scheme were applied (with queue $\sharp 1$ having highest priority). A similar   situation holds for queue $\sharp 2$ when load $\varrho_2$ decreases. This confirms the interest of the SQF discipline to favour traffic flows with least intensity.


\section{Analytic extensions}
\label{AE}


In this section, 
we extend the analyticity domain of functions $M_1$ and $M_2$ and determine their smallest singularities. Recall by Proposition \ref{Hexp} that they are analytic at least on the half-plane $\{z \in \mathbb{C} \mid \; \Re(z) > \max(\widetilde{s}_1,\widetilde{s}_2)/2\}$.

\subsection{Analytic continuation of function $\mathbf{M}$}
\label{ACM}

A property is said to hold \textit{generically} if it does for almost all $(\lambda_1, \lambda_2,\mu_1,\mu_2)$ in $\mathbb{R}^{4}$ with respect to Lebesgue measure.

\begin{theorem} 
Let $\Delta_j(z)$, $j \in \{1,2\}$, denote the discriminant of polynomial $R_j(w,z)$ in variable $w$, as introduced in Lemma \ref{Roots}.a.
\\
\indent
\textbf{a)} Discriminant $\Delta_j(z)$ has generically four distinct roots $\eta_j^{(1)}$, ..., $\eta_j^{(4)}$, two of those roots being real negative and the two others non real (complex conjugate). Let then $\eta_j^{(1)}$, $\eta_j^{(2)}$ denote the two real roots.
\\
\indent
\textbf{b)} 
Define the solution $\alpha_1(z)$ (resp. $\gamma_2(z)$) of equation $R_1(z,w) = 0$ (resp. $R_2(z,w) = 0$) as in Lemma \ref{Roots}.b. 
\\
\indent
Algebraic function $\alpha_1$ (resp. $h_1$) is analytic (resp. meromorphic) on $\widehat{\mathbb{C}} \setminus [\eta_1^{(1)},\eta_1^{(2)}]$. Symmetrically, algebraic function $\gamma_2$ (resp. $h_2$)  is analytic (resp. meromorphic) on $\widehat{\mathbb{C}} \setminus [\eta_2^{(1)},\eta_2^{(2)}]$.
\label{Delta12}
\end{theorem}
\begin{proof}
\textbf{a)} The localisation of the zeros of discriminant $\Delta_j(z)$ is detailed in Appendix \ref{A3bis}.I (for the existence of four distinct roots) and Appendix \ref{A3bis}.II (for the reality of two of them).
\\
\indent
\textbf{b)} Following Appendix \ref{A3bis}.I, equation $R_j(z,w) = 0$ equivalently represents the complex curve $\mathcal{R}_j$ in $\widehat{\mathbb{C}} \times \widehat{\mathbb{C}}$. As $R_j(z,w)$ has degree 3 in $w$, there consequently exists a 3-sheeted ramified covering $\Pi_j:w \in \mathcal{R}_j \mapsto z \in \widehat{\mathbb{C}}$ whose ramification points $z$ are either the roots of discriminant $\Delta_j(z)$ (determining multiple roots of $R_j(z,w)$) or possible points at infinity [\textbf{Fis01}, �9.6]. By Appendix \ref{A3bis}.I (Case I.B), there are no ramification at infinity and we conclude with \textbf{a)} that the only ramification points of $\Pi_j$ are the distinct roots $\eta_j^{(1)}$, ..., $\eta_j^{(4)}$ of $\Delta_j(z)$.
\\
\indent
By Lemma \ref{Roots}.b, $\alpha_j(z)$, $\beta_j(z)$ and $\gamma_j(z)$ are the roots of $R_j(z,w) = 0$, each of them being determined by inequalities for real $z > 0$. Each function $\epsilon_j \in \{\alpha_j,\beta_j,\gamma_j\}$ is then known to be meromorphic in $\widehat{\mathbb{C}}$ cut along segments $[\eta_j^{(1)},\eta_j^{(2)}]$ and $[\eta_j^{(3)},\eta_j^{(4)}]$ joining ramification points. Further, any function $\epsilon_j$ cannot take the value $\infty$ since the monomial in $w^3$ of $R_j(z,w)$ is non-zero by definition (\ref{R1}). We conclude that $\epsilon_j$ is actually analytic on $\widehat{\mathbb{C}} \setminus [\eta_j^{(1)},\eta_j^{(2)}] \cup [\eta_j^{(3)},\eta_j^{(4)}]$. 
\\
\indent
For given $z \in \widehat{\mathbb{C}}$, multiple solutions to equation $R_j(z,w) = 0$ have generically multiplicity 2. Moreover, it is easily verified through Cardano's formulae for solutions $\alpha_1(z)$, $\beta_1(z)$ and $\gamma_1(z)$ (see (\ref{Cardan}), Appendix \ref{A3bis}) that 
\begin{itemize}
\item $\alpha_1(z)$ and $\beta_1(z)$ coincide at real points $z = \eta_1^{(1)}$ and $z = \eta_1^{(2)}$;
\item $\beta_1(z)$ and $\gamma_1(z)$ coincide at non real points $z = \eta_1^{(3)}$ and $z = \eta_1^{(4)}$
\end{itemize}
and these solutions do not coincide otherwise. Symmetrically, solutions $\gamma_2(z)$ and $\beta_2(z)$ (resp. $\beta_2(z)$ and $\alpha_2(z)$) coincide for real points $z = \eta_2^{(1)}$ and $z = \eta_2^{(2)}$ only (resp. for non real points $z = \eta_2^{(3)}$ and $z = \eta_2^{(4)}$) and they do not coincide otherwise.
\\
\indent
By the previous discussion, it follows that function $\alpha_1$ (resp. $\gamma_2$) is actually analytic in the cut plane $\widehat{\mathbb{C}} \setminus [\eta_1^{(1)},\eta_1^{(2)}]$ (resp. $\widehat{\mathbb{C}} \setminus [\eta_2^{(1)},\eta_2^{(2)}]$). By definition (\ref{z2z2*}) of $h_1$, $h_1(z)$ is a rational function of both $z$ and $a_1(z) = z + \alpha_1(z)$; we then deduce that function $h_1$ is meromorphic on $\widehat{\mathbb{C}} \setminus [\eta_1^{(1)},\eta_1^{(2)}]$. Similarly, definition (\ref{z2z2**}) expresses $h_2(z)$ as a rational function of $z$ and $c_2(z) = z - \gamma_2(z)$, so that function $h_2$ is meromorphic on $\widehat{\mathbb{C}} \setminus [\eta_2^{(1)},\eta_2^{(2)}]$.
\end{proof}

We can now extend the analyticity domain of function $\mathbf{M} = (M_1,M_2)$ as follows.
\begin{prop}
Let $\tau_1 = \xi_1^-(\sigma_0^+)$ (resp. $\tau_2 = \xi_2^-(\sigma_0^+)$). 
\\
\indent
Function $\mathbf{M}$ can be analytically extended to the half-plane $\mathbf{V}_M$ defined by
\begin{itemize}
\item[\textbf{a1}.] $\mathbf{V}_M = \{z \in \mathbb{C} \; \mid \; \Re(z) > \frac{1}{2} \max(\sigma_0^+ + \tau_1,\sigma_0^+ + \tau_2) \}$ if ($I^+$) and ($II^+$) hold;
\item[\textbf{a2}.] $\mathbf{V}_M = \{z \in \mathbb{C} \; \mid \; \Re(z) > \max(\frac{1}{2}(\sigma_0^+ + \tau_1), \eta_2^{(1)} \}$ if ($I^-$) and ($II^+$) hold;
\item[\textbf{a3}.] $\mathbf{V}_M = \{z \in \mathbb{C} \; \mid \; \Re(z) > \max(\eta_1^{(1)}, \frac{1}{2} (\sigma_0^+ + \tau_2)) \}$ if ($I^+$) and ($II^-$) hold;
\item[\textbf{a4}.] $\mathbf{V}_M = \{z \in \mathbb{C} \; \mid \; \Re(z) > \max(\eta_1^{(1)},\eta_2^{(1)} \}$ if ($I^-$) and ($II^-$) hold,
\end{itemize}
where exclusive conditions ($I^+$), ($I^-$) and ($II^+$), ($II^-$) are stated in Lemma \ref{HoLpoles}. In the above defined domains $\mathbf{V}_M$, the smallest abcissa is  smaller than $\sigma_0^+$.
\label{extendM1M2}
\end{prop}
\begin{proof}
Solving linear system (\ref{G1G2-M1M2-}) for $M_1(z)$ and $M_2(z)$ readily gives
$$
\left\{
\begin{array}{ll}
M_1(z) = \displaystyle \frac{-\lambda_2\mu_2}{E(z)} \left[ \frac{(s_1-\xi_2^-(s_1))G_1(s_1)-J_1(s_1)}{\mu_2+s_2} + \frac{(s_2-\xi_1^-(s_2))G_2(s_2)-J_2(s_2)}{\mu_2 + \xi_2^-(s_1)}\right],
\nonumber \\ \\
M_2(z) = \displaystyle \frac{-\lambda_1\mu_1}{E(z)} \left[ \frac{(s_2-\xi_1^-(s_2))G_2(s_2)-J_2(s_2)}{\mu_1+s_1} + \frac{(s_1-\xi_2^-(s_1))G_1(s_1)-J_1(s_1)}{\mu_1 + \xi_1^-(s_2)}\right],
\end{array} \right.
$$
where
\begin{equation}
E(z) = \lambda_1\mu_1\lambda_2\mu_2 \left[ \frac{1}{(\mu_1+s_1)(\mu_2+s_2)} - \frac{1}{(\mu_1+\xi_1^-(s_2))(\mu_2+\xi_2^-(s_1))} \right]
\label{denE}
\end{equation}
and where $s_1$ and $s_2$ depend on variable $z$ according to $s_1 = C_2(z) = z + \gamma_2(z)$ and $s_2 = A_1(z) = z - \alpha_1(z)$, respectively. 
\\
\indent
As detailed in Appendix \ref{Psing}, Lemma \ref{lemmaE}, it can be first shown that denominator $E(z)$ cannot vanish for $\Re(z) > \max(\eta_1^{(1)},\eta_2^{(1)})$. Besides, we note that
\begin{itemize}
\item by Theorem \ref{Delta12}.b, $z \mapsto \alpha_1(z)$ (resp. $z \mapsto \gamma_2(z)$) is analytic on $\mathbb{C}$ cut along the segment joining its real ramification points, namely the real negative roots $\eta_1^{(1)}$, $\eta_1^{(2)}$ (resp. $\eta_2^{(1)}$, $\eta_2^{(2)}$) of discriminant $\Delta_1(z)$ (resp. $\Delta_2(z)$). We hereafter assume for instance that $\eta_1^{(2)} < \eta_1^{(1)} < 0$ (resp. $\eta_2^{(2)} < \eta_2^{(1)} < 0$); 
\item by Lemma \ref{Roots0}, $\xi_2^-$ (resp. $\xi_1^-$) is analytic on $\mathbb{C} \setminus [\zeta_1^-,\zeta_1^+]$ (resp. $\mathbb{C} \setminus [\zeta_2^-,\zeta_2^+]$);
\item by Corollary \ref{extensions}, $G_1$ (resp. $G_2$) is analytic on  $\widetilde{\omega}_1 = \{s_1 \in \mathbb{C} \mid \; \Re(s_1) > \widetilde{s}_1 \}$ (resp. $\widetilde{\omega}_2 = \{s_2 \in \mathbb{C} \mid \; \Re(s_2) > \widetilde{s}_2 \}$), with $\widetilde{s}_1$ and $\widetilde{s}_2$ defined in Lemma \ref{HoLpoles}. 
\end{itemize} 
From the expressions of $M_1(z)$ and $M_2(z)$ above and the latter properties, we deduce that $\mathbf{M}$ is analytic at any point $z$ such that $\Re(z) > \max(\eta_1^{(1)},\eta_2^{(1)})$ and
\begin{equation}
\Re(A_1(z)) > \max(\zeta_2^+,\widetilde{s}_2), \; \; \; \Re(C_2(z)) > \max(\zeta_1^+,\widetilde{s}_1).
\label{conditionsAC}
\end{equation}
According to which pair of conditions amongst $(I^+)$, $(II^+)$, $(I^-)$ and $(II^-)$ holds, the values in the right-hand sides of inequalities (\ref{conditionsAC}) are tabulated below:
$$
\begin{array}{|l|c|c|c|}
\hline \mathrm{\textbf{Case}} &  \max(\zeta_2^+,\widetilde{s}_2) & \max(\zeta_1^+,\widetilde{s}_1) \\[1 mm] 
\hline 
\textbf{a1.} \;  \; (I^+), (II^+) \;  & \sigma_0^+  & \sigma_0^+ \\[1 mm]
\textbf{a2.} \;  \; (I^-), (II^+) \;  & \sigma_0^+  & \zeta_1^+  \\[1 mm]
\textbf{a3.} \;  \; (I^+), (II^-) \;  & \zeta_2^+   & \sigma_0^+ \\[1 mm]
\textbf{a4.} \;  \; (I^-), (II^-) \;  & \zeta_2^+   & \zeta_1^+  \\[1 mm]
\hline
\end{array}
$$
\begin{figure}[b]
\scalebox{1}{\includegraphics[width=15cm, trim = 30 20 0 100,clip]{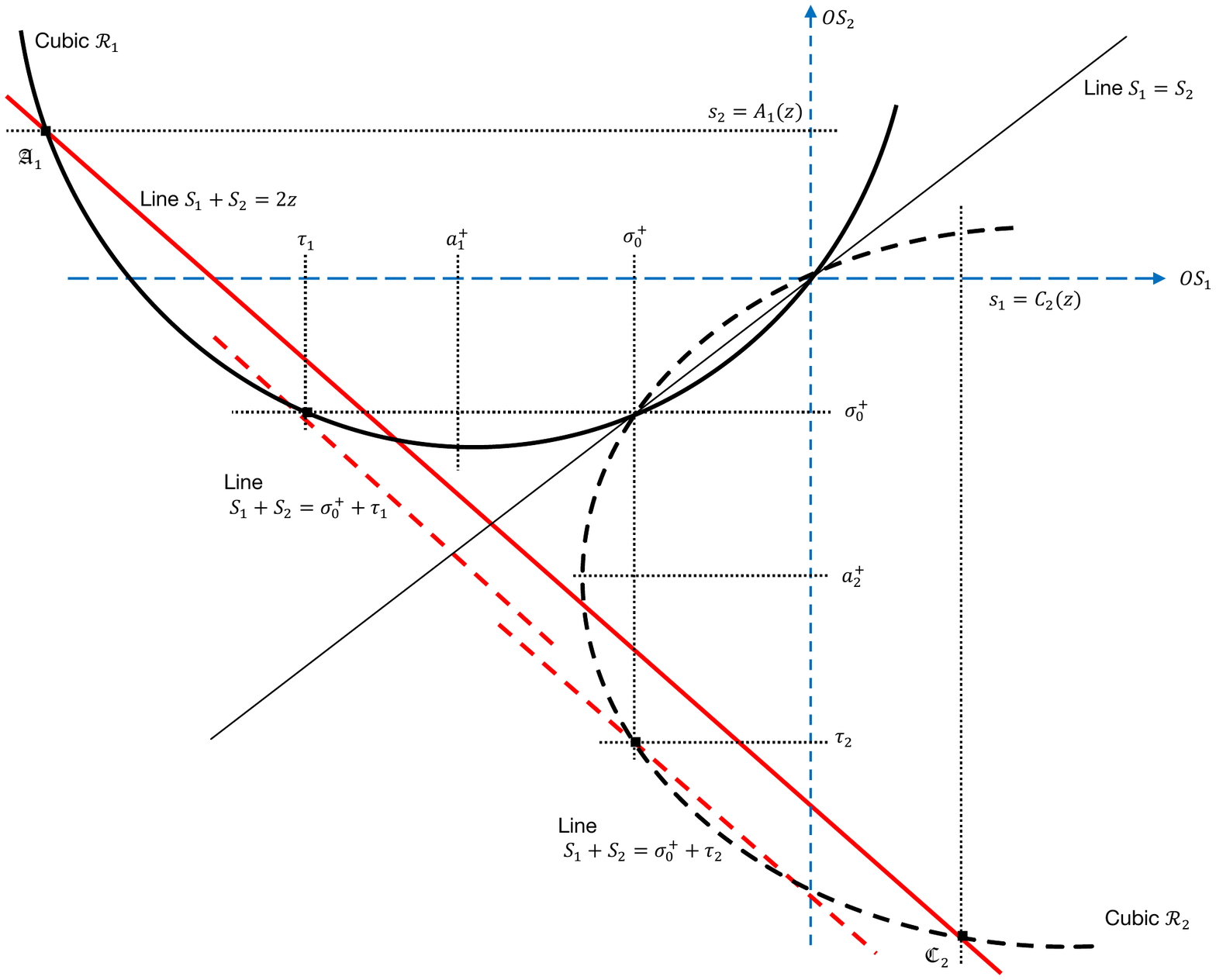}}
\caption{\textit{Configuration of cubics $\mathcal{R}_1$ and $\mathcal{R}_2$ in Case \textbf{a1} when conditions $(I^+)$ and $(II^+)$ hold.}}
\label{Fig6bis}
\end{figure}
\indent
Recall finally that argument $s_1 = C_2(z)$ (resp. $s_2 = A_1(z)$) is the abcissa (resp. the ordinate) of intersection point $\mathfrak{C}_2$ (resp. of intersection point $\mathfrak{A}_1$) of cubic $\mathcal{R}_2$ (resp. cubic $\mathcal{R}_1$) with line $S_1 + S_2 = 2z$ in the $(O,S_1,S_2)$ plane. Let us then successively consider the following cases:
\\
\indent 
\textbf{\textit{a1)}} \textbf{Case $(I^+)$ and $(II^+)$}. There exists a unique  ordinate $\tau_2 = \xi_2^-(\sigma_0^+)$ (resp. a unique abcissa $\tau_1 = \xi_1^-(\sigma_0^+)$) such that $T_2(\tau_2) = T_2(\sigma_0^+)  = \sigma_0^+$ (resp. $T_1(\tau_1) = T_1(\sigma_0^+)  = \sigma_0^+$)  with $\tau_2 < a_2^+ < \sigma_0^+$ and $\tau_1 < a_1^+ < \sigma_0^+$ (in fact, by the proof of Lemma \ref{HoLpoles}, condition $(I^+)$ ensures the existence of such a $\tau_2 < \sigma_0^+$, see Fig.\ref{Fig9}. Similarly, condition $(II^+)$ ensures the existence of such a $\tau_1 < \sigma_0^+$). 
\\
\indent
Now, the point $(\tau_1,\sigma_0^+) \in \mathcal{R}_1$ clearly pertains to the line $S_1 + S_2 = 2z_{0,1}$ with $2z_{0,1} = \tau_1 + \sigma_0^+$ (see Fig.\ref{Fig6bis}). By convexity of $T_1$ on interval $]\sigma_1^-,0[$ (\textbf{Remark \ref{Rem2}}, Appendix \ref{A3bis}), any line $S_1 + S_2 = 2z$ with $z > z_{0,1}$ cuts curve $\mathcal{R}_1$ at point $\mathfrak{A}_1$ with ordinate $A_1(z) > A_1(z_{0,1}) = \sigma_0^+$ and the first inequality (\ref{conditionsAC}) is ensured.
\\
\indent
Similarly, the point $(\sigma_0^+, \tau_2) \in \mathcal{R}_2$ belongs to the line $S_1 + S_2 = 2z_{0,2}$ with $2z_{0,2} = \tau_2 + \sigma_0^+$. Again by convexity of $T_2$ on interval $]\sigma_2^-,0[$, 
any line $S_1 + S_2 = 2z$ with $z > z_{0,2}$ cuts curve $\mathcal{R}_2$ at point $\mathfrak{C}_2$ with abcissa $C_2(z) > C_2(z_{0,2}) = \sigma_0^+$ and the second inequality (\ref{conditionsAC}) is ensured.
\\
\indent
The above discussion then implies that function $\mathbf{M} = (M_1,M_2)$ is analytic at least for $z > \max(z_{0,1},z_{0,2})$, hence for $\Re(z) > \max(z_{0,1},z_{0,2})$ (by definition (\ref{Hbis}), either $M_1$ or $M_2$ is the sum of two non-negative Laplace transforms).
\begin{figure}[b]
\scalebox{1}{\includegraphics[width=15cm, trim = 30 28 0 60,clip]{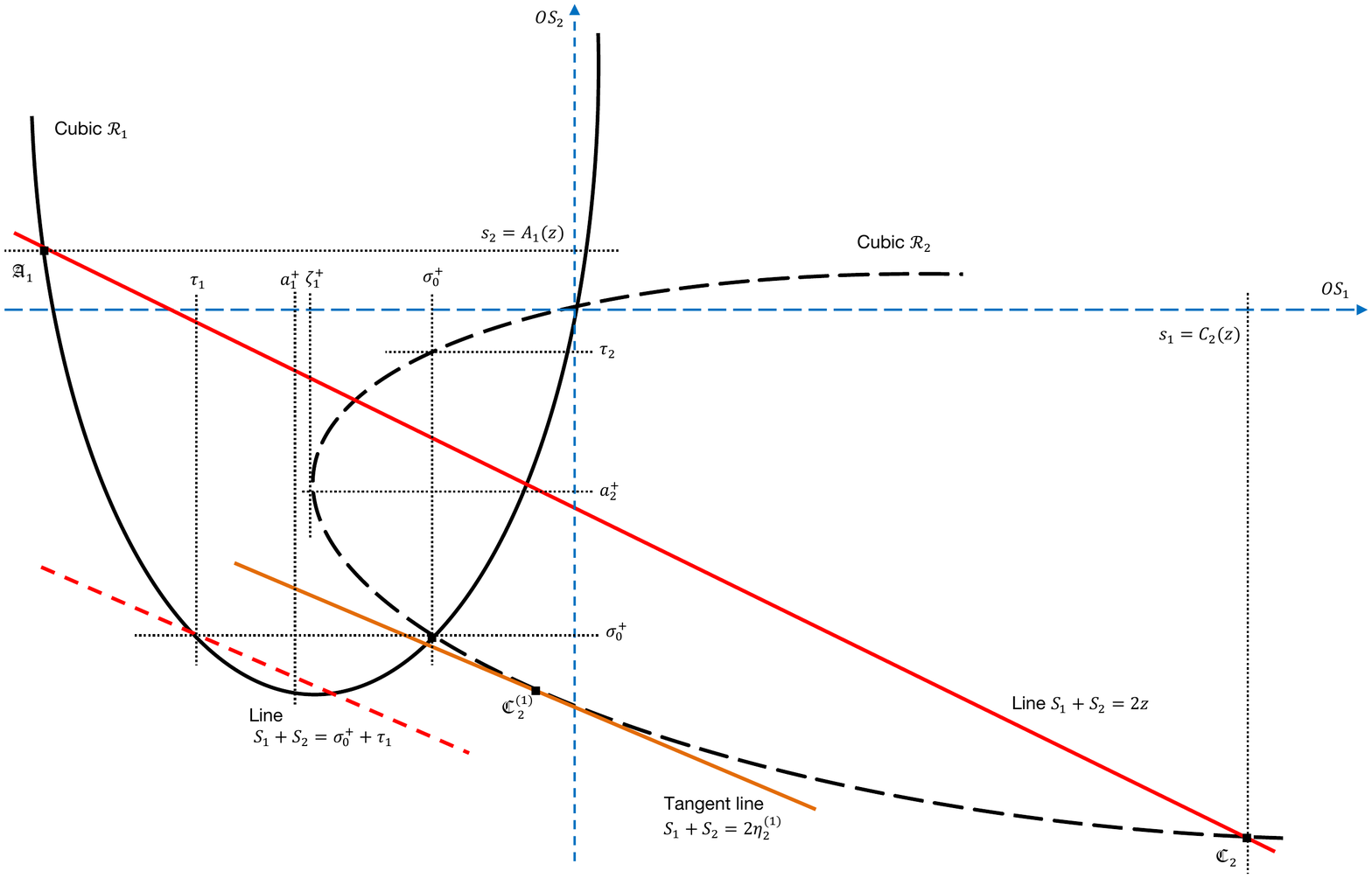}}
\caption{\textit{Configuration of cubics $\mathcal{R}_1$ and $\mathcal{R}_2$ in Case \textbf{a2} when conditions $(I^-)$ and $(II^+)$ hold.}}
\label{Fig6ter}
\end{figure}
\\
\indent
\textit{\textbf{a2)}} \textbf{Case $(I^-)$ and $(II^+)$}. Concerning curve $\mathcal{R}_1$, abcissa $\tau_1$ still verifies $\tau_1 < a_1^+ < \sigma_0^+$ by condition $(II^+)$. As in Case \textit{\textbf{a1}} above, we derive that $A_1(z) > \sigma_0^+$ for $z > z_{0,1}$, where $2z_{0,1} = \tau_1 + \sigma_0^+$ so that the first inequality (\ref{conditionsAC}) is ensured. 
\\
\indent
Concerning curve $\mathcal{R}_2$, ordinate $\tau_2$ now verifies $\sigma_0^+ < a_2^+ < \tau_2$ by condition $(I^-)$. Consider then the point $\mathfrak{C}_2^{(1)} = (T_2(c_2),c_2)$ where $T_2'(c_2) = -1$ (see Fig.\ref{Fig6ter}). From Appendix \ref{A3bis}.II (\textit{mutatis mutandis} from $\mathcal{R}_1$ to $\mathcal{R}_2$), point $\mathfrak{C}_2^{(1)}$ belongs to the tangent line $S_1 + S_2 = 2\eta_2^{(1)}$ to $\mathcal{R}_2$. By convexity of $T_2$ (\textbf{Remark \ref{Rem2}}, Appendix \ref{A3bis}), curve $\mathcal{R}_2$ is above that line; writing the abcissa of any point $\mathfrak{C}_2$ as $C_2(z) = T_2(s_2)$ for some $s_2 \in \; ]\sigma_2^-,a_2^+[$, we then have $C_2(z) = T_2(s_2) > T_2(a_2^+) = \zeta_1^+$ for all $z > \eta_2^{(1)}$ and the second inequality (\ref{conditionsAC}) is ensured. 
\\
\indent
The above discussion consequently shows that function $\textbf{M} = (M_1,M_2)$ is analytic at least for $z > \max(z_{0,1},\eta_2^{(1)})$, and thus for $\Re(z) > \max(z_{0,1},\eta_2^{(1)})$.
\\
\indent
\textbf{\textit{a3) Case $(I^+)$ and $(II^-)$}} or \textbf{\textit{a4) Case $(I^-)$ and $(II^-)$}}. An extended analyticity domain to function $\mathbf{M}$ is similarly derived for these cases on the basis of the respective configurations of cubics $\mathcal{R}_1$ and $\mathcal{R}_2$.
\end{proof}

\subsection{Smallest module singularities}

Corollary \ref{extensions} ensures that $G_1$ (resp. $G_2$) has no singularity in $\{s \in \mathbb{C} \; \mid \; \Re(s) > \widetilde{s}_1\}$ (resp. in $\{s \in \mathbb{C} \; \mid \; \Re(s) > \widetilde{s}_2\}$) where thresholds $\widetilde{s}_1$ and $\widetilde{s}_2$ are specified in Lemma \ref{HoLpoles}. Proposition \ref{extendM1M2} will now enable us to specify the smallest singularity of meromorphic transforms $G_1$ and $G_2$.
\begin{theorem}
Let constants
$$
\left\{
\begin{array}{ll}
r_{0,1} = \displaystyle \frac{T_2'(\sigma_0^+)}{T_2'(\sigma_0^+) - 1} \left [ J_1(\sigma_0^+) - \frac{\lambda_1\mu_1M_1(\sigma_0^+)}{\mu_1+\sigma_0^+} + \frac{\lambda_2\mu_2M_2(\sigma_0^+)}{\mu_2 + \sigma_0^+} \right ],
\\ \\
r_{0,2} = \displaystyle \frac{T_1'(\sigma_0^+)}{T_1'(\sigma_0^+) - 1} \left [ J_2(\sigma_0^+) + \frac{\lambda_1\mu_1M_1(\sigma_0^+)}{\mu_1+\sigma_0^+} - \frac{\lambda_2\mu_2M_2(\sigma_0^+)}{\mu_2 + \sigma_0^+} \right ],
\end{array} \right.
$$
where functions $J_1$ and $J_2$ are given by (\ref{J1J2}) (and $\psi_1(0)$ and $\psi_2(0)$ determined by Proposition \ref{psi12}.a), and
$$
\left\{
\begin{array}{ll}
r_1^+ = \displaystyle \frac{\sqrt{D_{0,1}(\zeta_1^+ - \zeta_1^-)}}{2(\mu_1 + \zeta_1^+)(\zeta_1^+-a_2^+)} \left [ G_1(\zeta_1^+) - \frac{\lambda_2\mu_2M_2(z_2^+)}{(\mu_2+a_2^+)^2} \right ],
\\ \\
r_2^+ = \displaystyle \frac{\sqrt{D_{0,2}(\zeta_2^+ - \zeta_2^-)}}{2(\mu_2 + \zeta_2^+)(\zeta_2^+-a_1^+)} \left [ G_2(\zeta_2^+) - \frac{\lambda_1\mu_1M_1(z_1^+)}{(\mu_1+a_1^+)^2} \right ]
\end{array} \right.
$$
with $D_{0,1} = 4\lambda_1\lambda_2 + (\mu_2+\lambda_1-\lambda_2)^2$, $z_2^+ = (\zeta_1^+ + a_2^+)/2$ and where $\zeta_1^+$, $\zeta_1^-$ are given in (\ref{zeta1+-}) (resp. $D_{0,2} = 4\lambda_1\lambda_2 + (\mu_1+\lambda_2-\lambda_1)^2$, $z_1^+ = (\zeta_2^+ + a_1^+)/2$ and $\zeta_2^+$, $\zeta_2^-$ obtained from (\ref{zeta1+-}) by permuting indexes 1 and 2).
\\
\indent
In \textbf{Cases a1, a2, a3} and \textbf{a4} of Proposition \ref{extendM1M2}, the singularities with smallest module of transforms $G_1$ and $G_2$ are defined by 
\\
\indent
\textbf{a1)} a simple pole at $s_1 = \sigma_0^+$ for $G_1$ (resp. a simple pole at $s_2 = \sigma_0^+$ for $G_2$) with residue $r_{0,1}$ (resp. residue $r_{0,2}$);
\\
\indent
\textbf{a2)} an algebraic singularity with order 1 at $s_1 = \zeta_1^+$ for $G_1$ (resp. a simple pole at $s_2 = \sigma_0^+$ for $G_2$) with residue $r_1^+$ (resp. residue $r_{0,2}$);
\\
\indent
\textbf{a3)} a simple pole at $s_1 = \sigma_0^+$ for $G_1$ (resp. an algebraic singularity with order 1 at $s_2 = \zeta_2^+$ for $G_2$) with residue $r_{0,1}$ (resp. residue $r_2^+$);
\\
\indent
\textbf{a4)} an algebraic singularity with order 1 at $s_1 = \zeta_1^+$ for $G_1$ (resp. an algebraic singularity with order 1 at $s_2 = \zeta_2^+$ for $G_2$) with residue $r_1^+$ (resp. residue $r_2^+$).
\label{Gsing12}
\end{theorem}
\begin{proof}
Consider the following cases:
\\
\indent
$\bullet$ \textit{\textbf{Case $(I^+)$}}. Write the 1st equation (\ref{G1G2-M1M2+}) as
\begin{equation}
G_1(s_1) = \displaystyle \frac{1}{s_1-\xi_2^+(s_1)} \left [ J_1(s_1) -  \frac{\lambda_1\mu_1M_1 (h_2(z))}{\mu_1+s_1}  + \frac{\lambda_2\mu_2M_2( h_2(z))}{\mu_2+\xi_2^+(s_1)}  \right ];
\label{G1-M1M2}
\end{equation}
as $s_1 \rightarrow \sigma_0^+$, we have $\xi_2^+(s_1) \rightarrow \sigma_0^+$ while $h_2(z) = (s_1 + \xi_2^+(s_1))/2 \rightarrow \sigma_0^+$. Following Proposition \ref{extendM1M2}.a1, functions $M_1 \circ h_2$ and $M_2 \circ h_2$ are analytic at $z = \sigma_0^+$ since $\frac{1}{2}\max(\sigma_0^+ + \tau_1,\sigma_0^+ + \tau_2) < \sigma_0^+$. By Corollary \ref{extensions}, $G_1(s_1)$ has presently no singularity for $\Re(s_1) > \widetilde{s}_1 = \sigma_0^+$; we then conclude from expression (\ref{G1-M1M2}) that $G_1$ has a simple pole at $s = \sigma_0^+$ with residue
$$
r_{0,1} = \frac{1}{1-\xi_2^+{'} (\sigma_0^+)} \left [ J_1(\sigma_0^+) - \frac{\lambda_1\mu_1M_1(\sigma_0^+)}{\mu_1+\sigma_0^+} + \frac{\lambda_2\mu_2M_2(\sigma_0^+)}{\mu_2 + \sigma_0^+} \right ].
$$
Using identity $T_2 \circ \xi_2^+ = \mathrm{Id}$ yields $\xi_2^{+}{'}(\sigma_0^+) = 1/T_2'(\sigma_0^+)$; residue $r_{0,1}$ then follows.
\\
\indent 
$\bullet$ \textit{\textbf{Case $(I^-)$}}. Letting $s_1  \rightarrow \sigma_0^+$, we presently have $\xi_2^+(s_1) \rightarrow \tau_2$ and therefore $h_2(z) = (s_1 + \xi_2^+(s_1))/2 \rightarrow z_{0,2}$ where $2z_{0,2} = \sigma_0^+ + \tau_2$. Proposition \ref{extendM1M2} then ensures that $M_1 \circ h_2$ and $M_2 \circ h_2$ are analytic at $z = z_{0,2}$ since $z_{0,2} > \max(z_{0,1},\eta_2^{(1)})$ (in fact, we clearly have $z_{0,2} > \eta_2^{(1)}$, as chord $S_1 + S_2 = 2z_{0,2}$ is  above tangent $S_1 + S_2 = 2\eta_2^{(1)}$; besides, our assumption $\tau_1 < \sigma_0^+ < \tau_2$ implies $z_{0,2} > z_{0,1}$). We conclude from (\ref{G1-M1M2}) and the latter discussion that $\sigma_0^+$ is not a singularity of $G_1$. Furthermore, 2nd condition (\ref{conditionsAC}) ensures that $M_1$ and $M_2$ are analytic at any point $z$ for which $\Re(C_2(z)) > \max(\zeta_1^+,\widetilde{s}_1) = \zeta_1^+$; by (\ref{G1-M1M2}) again, we conclude that $G_1$ is analytic at any point $s_1$ for which $\Re(s_1) > \zeta_1^+$. 
\\
\indent
To specify the nature of point $\zeta_1^+$ for $G_1$, use then formula (\ref{xi2+-}) for $\xi_2^+(s_1)$, where discriminant $D_1(s_1)$ is written as $D_1(s_1) = D_{0,1}(s_1-\zeta_1^-)(s_1-\zeta_1^+)$ with $D_{0,1} = 4\lambda_1\lambda_2 + (\mu_2+\lambda_1-\lambda_2)^2$; we obtain
$$
\xi_2^+(s_1) = a_2^+ + E_{0,1}\sqrt{s_1-\zeta_1^+} + o(s_1-\zeta_1^+)^{1/2}
$$
where $a_2^+$ is the largest stationary point of $T_2$ given by (\ref{statpoints2}) and with constant $E_{0,1} = [D_{0,1}(\zeta_1^+-\zeta_1^-)]^{1/2}/2(\mu_1+\zeta_1^+)$. Besides, $z = (s_1+\xi_2^-(s_1))/2$ tends to $z_2^+ = (\zeta_1^+ + a_2^+)/2$ as $s_1 \rightarrow \zeta_1^+$. By (\ref{G1-M1M2}) and the latter expansions, we then obtain
\begin{align}
G_1(s_1) = \frac{1}{\zeta_1^+ - a_2^+ - E_{0,1}\sqrt{t}+o(\sqrt{t})} \Big [ & J_1(\zeta_1^+) - \frac{\lambda_1\mu_1M_1(z_2^+)}{\mu_1+\zeta_1^+} +
\nonumber \\
& \frac{\lambda_2\mu_2M_2(z_2^+)}{\mu_2+a_2^++E_{0,1}\sqrt{t}+o(\sqrt{t}) } + o(\sqrt{t}) \Big ]
\nonumber 
\end{align}
after some simple algebra, with $t = s_1-\zeta_1^+$ for short; this provides us the final expansion $G_1(s_1) = G_1(\zeta_1^+) + r_1^+ \sqrt{t} + o(\sqrt{t})$ with
$$
r_1^+ = \frac{E_{0,1}}{\zeta_1^+-a_2^+} \left [ G_1(\zeta_1^+) - \frac{\lambda_2\mu_2M_2(z_2^+)}{(\mu_2+a_2^+)^2} \right ]
$$
and where $E_{0,1}$ is expressed as above. We conclude that the singularity with smallest module of $G_1$ is $\zeta_1^+$, an algebraic singularity with order 1 and residue $r_1^+$.
\\
\indent
Cases $(II^+)$ and $(II^-)$ for transform $G_2$ are similarly treated, \textit{mutatis mutandis}. Mixed cases \textbf{\textit{a1}}, \textbf{\textit{a2}}, \textbf{\textit{a3}} and \textbf{\textit{a4}} are then readily derived from the above discussion. 
\end{proof}
\begin{figure}[t]
\scalebox{1}{\includegraphics[width=14cm, trim = 30 100 0 60,clip]{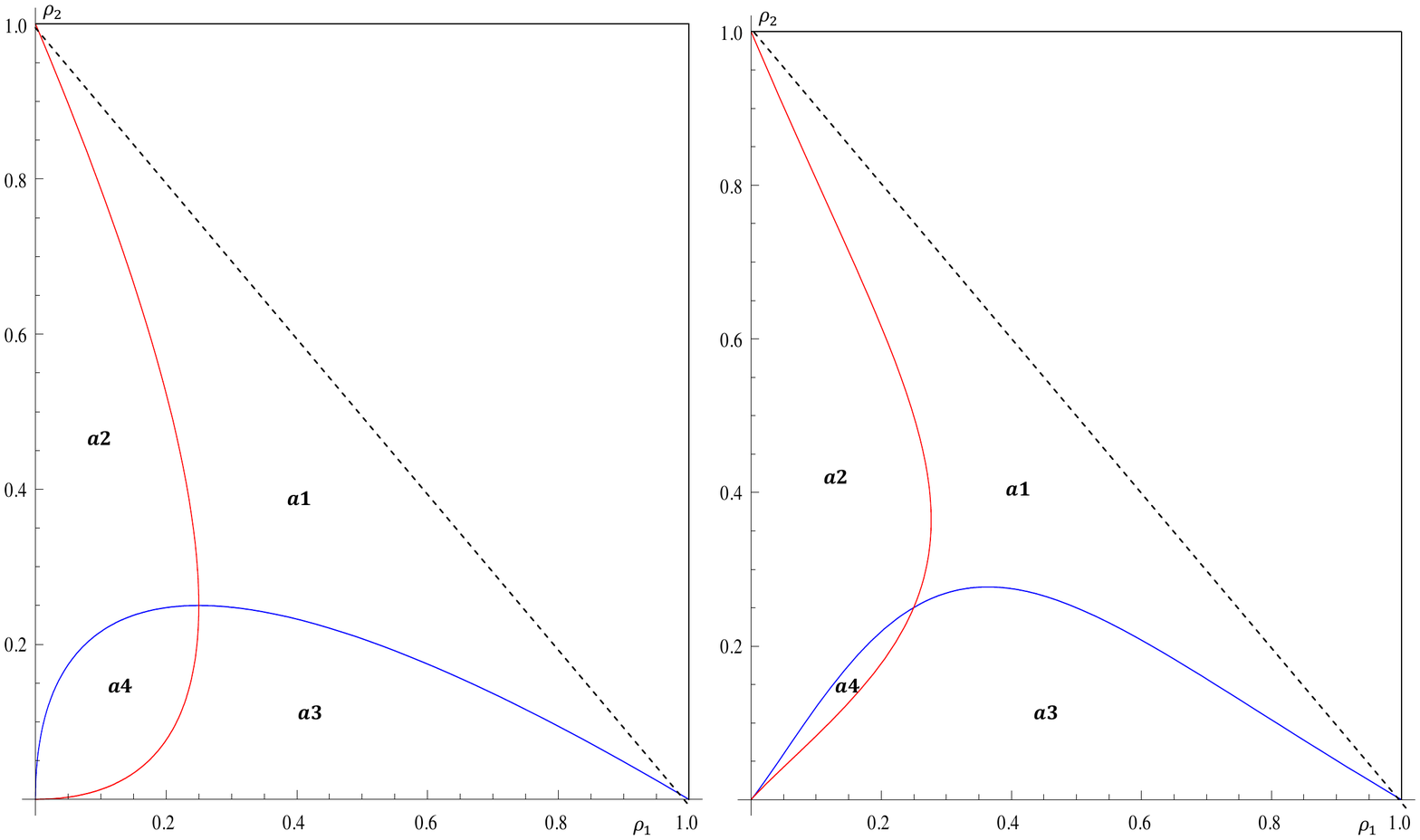}}
\caption{\textit{Regions of the $(O,\varrho_1,\varrho_2)$-plane associated with  \textbf{Cases} \textbf{a1}, \textbf{a2}, \textbf{a3},  \textbf{a4} for $\mu_1 = \mu_2$ (left) and $\lambda_1 = \lambda_2$ (right).}}
\label{Fig6quater}
\end{figure}
As an example, \textbf{Case} \textbf{\textit{a3}} holds for the values $\lambda_1 = 1$, $\lambda_2 = 1$, $\mu_1 = 3$, $\mu_2 = 4.5$ considered in Fig.\ref{Fig4}, where  $\widetilde{s}_1 = \sigma_0^+ \approx -1.5$ and $\widetilde{s}_2 = \zeta_2^+ \approx -1.57$. More generally, we represent in Fig.\ref{Fig6quater} the respective regions of the $(O,\varrho_1,\varrho_2)$ plane associated with \textbf{Cases} \textbf{\textit{a1}}, \textbf{\textit{a2}}, \textbf{\textit{a3}} and \textbf{\textit{a4}} in two specific situations: 
\begin{itemize}
\item[(1)] $\mu_1 = \mu_2$, for which condition $(I^+)$ (resp. condition $(II^+)$) reads $\varrho_1 > f(\varrho_2)$ (resp. $\varrho_2 > f(\varrho_1)$) where we set $
f(x) = \sqrt{x}(1-\sqrt{x})$, $0 \leq x \leq 1$;
\item[(2)] $\lambda_1 = \lambda_2$, for which condition $(I^+)$ (resp. condition $(II^+)$) reads $\varrho_1 > g(\varrho_2)$ (resp. $\varrho_2 > g(\varrho_1)$) where we set $g(x) = x(1-\sqrt{x})/(1+2x-2\sqrt{x})$, $0 \leq x \leq 1$.
\end{itemize}


\section{Large queue asymptotics}
\label{LQA}


We finally address the derivation of asymptotics for the distribution of workloads $U_1$ and $U_2$, that is, estimates of tail probabilities $\mathbb{P}(U_1 > u_1)$ and $\mathbb{P}(U_2 > u_2)$ for large queue content $u_1$ and $u_2$. We recall the following Tauberian theorem relating the singularities of a Laplace transform to the asymptotic behaviour of its inverse.
\begin{theorem} [\textbf{Doe58}, Theorem 25.2, p.237] Let $F$ be a Laplace transform and $\omega$ be its singularity with smallest module, with  
$F(s) \sim \kappa_0 (s - \omega)^{\nu_0}$ as $s \rightarrow \omega$ 
for $\kappa_0 \neq 0$ and $\nu_0 \notin \mathbb{N}$ (replace $F$ by $F - F(\omega)$ if $F(\omega)$ is finite). The Laplace inverse $f$ of $F$ is then estimated by 
$$
f(u) \sim \frac{\kappa_0}{\Gamma(-\nu_0)} \frac{ e^{\omega u}}{u^{\nu_0 + 1}}
$$
for $u \uparrow +\infty$, where $\Gamma$ denotes Euler's Gamma function.
\label{Tauberian}
\end{theorem}


\subsection{Rough estimates and the relation with the HoL policy}
\label{GS}

Following upper bound $U_1 \leq \overline{U}_1$ relating $U_1$ to variable $\overline{U}_1$ corresponding to a HoL service policy with highest priority given to queue $\sharp 2$ ([\textbf{Gui12}], Proposition 3.2), we have 
\begin{equation}
\mathbb{P}(U_1 > u_1) \leq \mathbb{P}(\overline{U}_1 > u_1)
\label{Rough1}
\end{equation}
for all $u_1 \geq 0$. By Lemma \ref{HoLpoles}, the Laplace transform $\overline{F}_1$ of $\overline{U}_1$ is meromorphic on the cut plane $\mathbb{C} \setminus [\zeta_1^-,\zeta_1^+]$, with a possible pole at the zero $\sigma_0^+$ of $P$ if some inequality on system parameters is fulfilled. Specifically, the application of Theorem \ref{Tauberian} shows that the tail behaviour of $\overline{U}_1$ is given by
\begin{equation}
\mathbb{P}(\overline{U}_1 > u_1) = 
\left\{
\begin{array}{lll}
O(e^{\sigma_0^+ u_1}) & \; \mbox{ if } \mathrm{(}I^+\mathrm{)} \; \mbox{holds},
\\
\\
\displaystyle O\left(\frac{e^{\zeta_1^+ u_1}}{\sqrt{u_1}}\right) & \; \mbox{ if } \mathrm{(}I^-\mathrm{)} \; \mbox{holds}
\end{array} \right.
\label{asymptHoL}
\end{equation} 
for large $u_1$, where conditions ($I^+$) and ($I^-$) have been stated in Lemma \ref{HoLpoles}. The tail behaviour of $\overline{U}_1$, and therefore $U_1$, may therefore be either exponential or subexponential according to system parameters. A similar discussion provides the tail behaviour of $U_2$ when compared to variable $\overline{U}_2$ corresponding to a HoL policy with highest priority given to queue $\sharp 1$.


\subsection{Tail asymptotics}


Let us now determine precise asymptotics for the tail behaviour of either variable $U_1$ or $U_2$. Applying definition (\ref{Ldef}) to $s_2 = 0$ gives the Laplace transform of $U_1$ as $F(s_1,0) = 1 - \varrho + H_1(s_1,0) + H_2(s_1,0)$, that is,
\begin{equation}
F(s_1,0) = 1 - \varrho + F_1(s_1,0) + G_1(s_1) + F_2(s_1,0) + G_2(0)
\label{LaplU1}
\end{equation}
with 
\begin{align}
\left \{
\begin{array}{ll}
F_1(s_1,0) & = \displaystyle \frac{J_2(0)-K_2(s_1,0)G_2(0)}{K_1(s_1,0)} \; \; \; +  \displaystyle \frac{H(s_1,0)}{K_1(s_1,0)},
\nonumber \\ \\
F_2(s_1,0) & = \displaystyle \frac{J_1(s_1)-K_1(s_1,0)G_1(s_1)}{K_2(s_1,0)} - \displaystyle \frac{H(s_1,0)}{K_2(s_1,0)},
\nonumber \\ \\
H(s_1,0) & = \displaystyle \frac{\lambda_1\mu_1}{\mu_1+s_1}M_1\left( \frac{s_1}{2} \right) - \lambda_2 M_2\left( \frac{s_1}{2} \right)
\end{array} \right.
\end{align}
after expressions (\ref{f1f2bis}) and (\ref{Hbis}), respectively.
\begin{theorem}
For large $u_1$, we have
\begin{equation}
\mathbb{P}(U_1 > u_1) \sim 
\left\{
\begin{array}{lll}
\displaystyle - \frac{(\sigma_0^+ + \mu_1)r_{0,1}}{\lambda_1\sigma_0^+} \cdot e^{\sigma_0^+ u_1} & \; \mbox{ if } \mathrm{(}I^+\mathrm{)} \; \mbox{holds},
\\
\\
\displaystyle \frac{(\zeta_1^+ + \mu_1)r_1^+}{2\lambda_1\zeta_1^+ \sqrt{\pi}} \cdot \frac{e^{\zeta_1^+ u_1}}{u_1^{3/2}} & \; \mbox{ if } \mathrm{(}I^-\mathrm{)} \; \mbox{holds}
\end{array} \right.
\label{asymptSQF}
\end{equation}
with residues $r_{0,1}$ and $r_1^+$ introduced in Theorem \ref{Gsing12}.
\\
\indent
Asymptotics of $\mathbb{P}(U_2 > u_2)$ for large $u_2$ is similarly derived according as either condition $(II^+)$ or $(II^-)$ holds.
\label{Asymptotics}
\end{theorem} 
\begin{proof}
Assume first that $(I^+)$ holds. By Proposition \ref{extendM1M2}, the expression (\ref{LaplU1}) of $H(s_1,0)$ in terms of $M_1$ and $M_2$ shows that $s_1 \mapsto H(s_1,0)$ is analytic for $\Re(s_1) >  \max(\sigma_0^++\tau_1,\sigma_0^++\tau_2)$ and $\Re(s_1) > \max(2\eta_1^{(1)},\sigma_0^++\tau_2)$, both conditions encompassing point $s_1 = \sigma_0^+$. We then deduce from (\ref{LaplU1}) that the singularity with smallest module of $F(s_1,0)$ is at $s_1 = \sigma_0^+$ with leading term
\begin{equation}
F(s_1,0) \sim - \frac{K_2(s_1,0)}{K_1(s_1,0)}G_1(s_1) + G_1(s_1) = \frac{s_1 + \mu_1}{\lambda_1}G_1(s_1)
\label{AsymptU1}
\end{equation}
since by definition
$$
\frac{K_1(s_1,0)}{K_2(s_1,0)} = - \frac{s_1+\mu_1-\lambda_1}{\lambda_1}
$$
and the root $\lambda_1 - \mu_1$ of $K_1(s_1,0)$ is less than $\sigma_0^+$ (in fact,  $P(\lambda_1 - \mu_1) = -\lambda_1\lambda_2 < 0$ by (\ref{PolP})). By Theorem \ref{Gsing12}, $s_1 = \sigma_0^+$ is a simple pole for $G_1$ with residue $r_{0,1}$; applying Tauberian Theorem \ref{Tauberian} to asymptotics (\ref{AsymptU1}) with $\kappa_0 = (\sigma_0^+ + \mu_1)r_{0,1}/\lambda_1$ and $\nu_0 = -1$, we derive that $\mathbb{P}(U_1 > u) \sim -(\sigma_0^+ + \mu_1)r_{0,1}e^{\sigma_0^+ u}/\lambda_1\sigma_0^+$ for large $u$, as announced.
\\
\indent
Assume now that $(I^-)$ holds. It is readily verified that the line $S_1 + S_2 = Const.$ passing by $(\zeta_1^+,a_2^+)$ is above the parallel line passing by $(\sigma_0^+,\sigma_0^+)$ (see Fig.\ref{Fig6ter}); as $\sigma_0^+$ is also larger than either $\eta_1^{(1)}$ or $\eta_2^{(1)}$, we then derive that
$$
\frac{\zeta_1^+ + a_2^+}{2} > \sigma_0^+ > \max(\eta_1^{(1)},\eta_2^{(1)})
$$
hence $\zeta_1^+/2 > \sigma_0^+ > \max(\eta_1^{(1)},\eta_2^{(1)})$ since  $a_2^+ < 0$. By Proposition \ref{extendM1M2} and the expression (\ref{LaplU1}) of $H(s_1,0)$ in terms of $M_1$ and $M_2$, the latter inequalities ensure that function $s_1 \mapsto H(s_1,0)$ is analytic at $s_1 = \zeta_1^+$. It then follows from (\ref{LaplU1}) that the singularity with smallest module of $F(s_1,0)$ is at $s_1 = \zeta_1^+$ with leading term provided by (\ref{AsymptU1}) again, so that
\begin{equation}
F(s_1,0) - F(\zeta_1^+,0) \sim \frac{\zeta_1^+ + \mu_1}{\lambda_1} \left [ G_1(s_1) - G_1(\zeta_1^+) \right ]
\label{AsymptU1bis}
\end{equation}
near $s_1 = \zeta_1^+$. By Theorem \ref{Gsing12}, $s_1 = \zeta_1^+$ is an algebraic singularity for $G_1$ with residue $r_1^+$; (\ref{AsymptU1bis}) then gives $F(s_1,0) - F(\zeta^+,0) \sim r_1 (s_1-\zeta_1^+)^{1/2}$ as $s_1 \rightarrow \zeta_1^+$ where $r_1 = (\zeta_1^+ + \mu_1)r_1^+/\lambda_1$. 
Applying Tauberian Theorem \ref{Tauberian} with $\kappa_0 = r_1$, $\nu_0 = 1/2$, we then derive that $\mathbb{P}(U_1 > u) \sim \kappa_1 e^{\zeta_1^+ u}/u^{3/2}$ for large $u$ with prefactor $\kappa_1 = -r_1/\zeta_1^+\Gamma(-1/2) = r_1/2\zeta_1^+ \sqrt{\pi}$, as claimed.
\end{proof}
Theorem \ref{Asymptotics} eventually specifies the tail behaviour for the respective distributions of workloads $U_1$ and $U_2)$, depending on parameter values $\lambda_1$, $\mu_1$, $\lambda_2$, $\mu_2$ with $\varrho_1 + \varrho_2 < 1$; in a summarized form, we have shown that
$$
\mathbb{P}(U_1 > u_1) = O(e^{\widetilde{s}_1u_1}), \; \; \; \mathbb{P}(U_2 > u_2) = O(e^{\widetilde{s}_2u_2})
$$
with rates $\widetilde{s}_1$ and $\widetilde{s}_2$ given by Lemma \ref{HoLpoles}.  Specifically, Case \textbf{\textit{a1}} gives exponential decay at infinity to both queues with identical rate $\sigma_0^+$, while last Case \textbf{\textit{a4}} corresponds to subexponential decays  with respective rate $\zeta_1^+$ and $\zeta_2^+$; finally, Case \textbf{\textit{a2}} and Case \textbf{\textit{a3}} correspond to mixed exponential/subexponential behaviours.
\\
\indent
For illustration, assume that mean service times have identical mean ($\mu_1 = \mu_2$) and that queue $\sharp 1$ receives low intensity traffic, i.e. $\varrho_1$ tends to 0. By Fig.\ref{Fig6quater}, this corresponds to either 
\begin{itemize}
\item Case \textbf{\textit{a2}} for which $\mathbb{P}(U_1 > u_1) = O(e^{\zeta_1^+u_1})$ and $\mathbb{P}(U_2 > u_2) = O(e^{\sigma_0^+u_2})$ with $-\zeta_1^+ > -\sigma_0^+$, so that queue $\sharp 1$ is smaller than queue $\sharp 2$ regarding the sharpness of distribution tails;
\item or Case \textbf{\textit{a4}} for which $\mathbb{P}(U_1 > u_1) = O(e^{\zeta_1^+u_1})$ and $\mathbb{P}(U_2 > u_2) = O(e^{\zeta_2^+u_2})$. By formulae (\ref{zeta1+-}), supplementary condition $\zeta_1^+ < \zeta_2^+$ easily reduces to $\varrho_1 < \varrho_2$, which is clearly fulfilled for small $\varrho_1$;  queue $\sharp 1$ is then smaller than queue $\sharp 2$ in the same sense.
\end{itemize}
In any of the latter cases, the dynamic SQF discipline consequently provides priority to the queue with less traffic intensity, as motivated by its definition.


\section{Conclusion}


As a generalisation to the static HoL priority scheme, the SQF discipline provides a dynamic scheme for controlling traffic congestion in favour of less congested queues. Within the Markovian framework, its mathematical analysis involves quite a challenging new setting, namely functional equations whose solutions expand as a  series involving the semi-group $<h_1,h_2 >$ generated by two algebraic functions $h_1$ and $h_2$; such functions prove to be naturally attached to a pair of rational cubics. As a result, the resolution framework developed in this paper has enabled us to derive main performance characteristics. 

To our knowledge, such functional equations for coupled queues have not appeared so far in the queueing theory literature. In the prior treatment of the symmetric case [\textbf{Gui12}], it was indicated that the analysis can be formulated as a Riemann-Hilbert problem; in the general asymmetric case, it can also be argued that a formulation as a two-dimensional Riemann-Hilbert problem is possible, where two independent boundary conditions hold for functions $M_1$ and $M_2$; note that such boundary conditions are not valid on closed contours but open arcs.

The analytic extension of $\mathbf{M} = (M_1,M_2)$ to some half-plane $\mathbf{V}_M \subset \mathbb{C}$ has also been shown to play a central role in the derivation of asymptotics for distribution tails. Besides, the convergence of the series expansion $\mathbf{M}(z)$ in terms of semi-group $<h_1,h_2>$ has been established for real positive values of $z$ only which, nevertheless, proves sufficient for the derivation of empty queue probabilities. It can be actually shown that such a series expansion converges uniformly for all complex $z$ pertaining to some subset of the form 
$\mathbf{W}_M = \{z \in \mathbb{C} \; \vert \; 2c - \Re(z) < \vert \Im(z) \vert \}$
for some constant $c$, therefore providing a meromorphic continuation of  $\mathbf{M}$ to $\mathbf{W}_M$. This extended meromorphic domain $\mathbf{W}_M$, clearly distinct from analyticity domain $\mathbf{V}_M$, can be derived as a subset of the so-called "Fatou set" associated with semi-group $<h_1,h_2>$, when the latter is seen as a holomorphic dynamical system \textbf{[Mil06]}. Compared to the classical theory dealing with semi-groups $<f>$, say, generated by a single holomorphic function $f$, our situation is new in that semi-group $<h_1,h_2>$ is generated by two independent holomorphic functions. Such developments can be envisaged for further investigation.

The analysis of other tightly coupled queueing systems could also be the object of further studies, e.g. the "Longest Queue First" LQF comparable discipline. Note that for such disciplines, the joint distribution could also involve other subsets of the state space which could not intervene for SQF, for instance the positive diagonal $u_1 = u_2 > 0$. This could certainly modify some aspects of the analysis but it is believed the presently developed resolution framework applies.


\section{Appendix}
\label{APP}



\subsection{Analytic continuation of $\xi_2^\pm$ and $\xi_1^\pm$}
\label{A30}


By expression (\ref{xi2+-}) and the fact that $\sqrt{D_1(-\mu_1)} = +\lambda_1\mu_1$, function $\xi_2^+$ is well defined on $\R \setminus [\zeta_1^-,\zeta_1^+]$ while function $\xi_2^-$ is defined on $\R \setminus ([\zeta_1^-,\zeta_1^+] \cup \{-\mu_1\}$ with $\xi_2^-(-\mu_1) = \infty$. In the following, we examine how functions $\xi_2^+$ and $\xi_2^-$ can be analytically continued to the cut plane $\C\setminus [\zeta_1^-,\zeta_1^+]$. 
\\
\indent
First note that for $\Re(s_1) = (\zeta_1^+ + \zeta_1^-)/2 <0$, we have $\Im(D_1(s_1)) = 0$ and $\Re(D_1(s_1)) < 0$ (note taht vertical line $\Re(s_1) =  (\zeta_1^+ + \zeta_1^-)/2$ and the real line $\Im(s_1) = 0$ are the only subsets of the complex plane on which $\Im(D_1(s_1)) = 0$). The Schwarz's reflection principle applied to function $\sqrt{D_1}$ with respect to the vertical line $\Re(s_1) =  (\zeta_1^+ + \zeta_1^-)/2$ then ensures that function $E_2$ defined by
$$
E_2(s_1) = 
\left\{ 
\begin{array}{ll}
-\sqrt{D_1(s_1)} \; \; \mathrm{if} \; \; \Re(s_1) \leq \displaystyle \frac{\zeta_1^+ + \zeta_1^-}{2},
\\ \\
+\sqrt{D_1(s_1)} \; \; \mathrm{if} \; \; \Re(s_1) \geq \displaystyle \frac{\zeta_1^+ + \zeta_1^-}{2},
\end{array}
\right.
$$
is globally analytic on the cut plane $\C\setminus[\zeta_1^+,\zeta_1^-]$. Define then functions $\xi_2$ and $\widetilde{\xi}_2$ in $\C\setminus [\zeta_1^-,\zeta_1^+]$ by
$$
\xi_2(s_1) = 
\left\{ 
\begin{array}{ll} 
\xi_2^-(s_1) \; \; \mathrm{if} \; \; \displaystyle \Re(s_1) \leq \frac{\zeta_1^+ + \zeta_1^-}{2}, \\ \\ 
\xi_2^+(s_1) \; \; \mathrm{if} \; \; \displaystyle \Re(s_1) \geq \frac{\zeta_1^+ + \zeta_1^-}{2},
\end{array}
\right.
\widetilde{\xi}_2(s_1) = 
\left\{ 
\begin{array}{ll}  
\xi_2^+(s_1) \; \; \mbox{if} \; \; \displaystyle \Re(s_1) \leq \frac{\zeta_1^+ + \zeta_1^-}{2}, \\ \\ 
\xi_2^-(s_1) \; \; \mbox{if} \; \; \displaystyle \Re(s_1) \geq \frac{\zeta_1^+ + \zeta_1^-}{2},
\end{array} 
\right. 
$$
respectively. By the analyticity of function $E_2$, it then follows that function $\xi_2$ is a meromorphic extension of $\xi_2^+$ to $\C\setminus[\zeta_1^+,\zeta_1^-]$ with a pole at point $-\mu_1$; besides, function $\widetilde{\xi}_2$ is an analytic extension of $\xi_2^-$ in $\C\setminus[\zeta_1^+,\zeta_1^-]$. To simplify notation, we will still refer to $\xi_2^+$ (resp. $\xi_2^-$) to denote its global meromorphic/analytic extension $\xi_2$ (resp. $\widetilde{\xi}_2$) to $\C\setminus[\zeta_1^+,\zeta_1^-]$.
\\
\indent
A similar reasoning applies to the meromorphic/analytic extension of functions $\xi_1^+$ and $\xi_1^-$ on the cut plane $\mathbb{C} \setminus [\zeta_2^-,\zeta_2^+]$, respectively.


\subsection{Proof of Lemma \ref{Roots}}
\label{A3}


\indent \textbf{a)} Using expression (\ref{Kexp}) for $K(s_1,s_2)$ and definition (\ref{Ker}), coefficient  $K_1(z+w,z-w) = K_1(s_1,s_2)$ reduces to
\begin{equation}
K_1(z+w,z-w) = - \frac{r_1(s_1,s_2)}{(s_1+\mu_1)(s_2+\mu_2)}
\label{K1z1z2}
\end{equation}
where $r_1(s_1,s_2) = -s_1^2s_2 + (\lambda - \mu_1)s_1s_2 - \mu_2s_1^2 - \mu_2(\mu_1-\lambda_1)s_1 + \lambda_2\mu_1 s_2$. Variable change (\ref{varch}) then gives $r_1(s_1,s_2) = R_1(w,z)$ with cubic polynomial $R_1(w,z)$ defined as in (\ref{R1}); expression (\ref{abexp}) for $K_1(z+w,z-w)$ follows. A similar calculation reduces coefficient $K_2(z+w,z-w) = K_2(s_1,s_2)$ to
$$
K_2(z+w,z-w) = \frac{r_2(s_1,s_2)}{(s_1+\mu_1)(s_2+\mu_2)}
$$
where $r_2(s_1,s_2) = s_1s_2^2 - (\lambda - \mu_2)s_1s_2 + \mu_1s_2^2 - \lambda_1\mu_2s_1 + \mu_1(\mu_2-\lambda_2)s_2$. Variable change (\ref{varch}) then gives $r_2(s_1,s_2) = R_2(w,z)$ with cubic polynomial $R_2(w,z)$ given as in (\ref{R2}). Expression (\ref{abexp}) for $K_2(z+w,z-w)$ follows (note that the opposite signs for $K_1(z+w,z-w)$ and $K_2(z+w,z-w)$ come from the fact that exchanging queue indexes $\sharp 1$ and $\sharp 2$ transforms $(w,z)$ into $(-w,z)$). 
\\
\indent
Finally, write
$$
K_2(s_1,s_2) - K_1(s_1,s_2) = s_2 - s_1 = -2w
$$
by definitions (\ref{Ker}) and (\ref{varch}); by the latter relation together with expressions (\ref{abexp}) of $K_1(z+w,z-w)$ and $K_2(z+w,z-w)$, we obtain identity (\ref{R1R2}) for the sum $R_1(w,z) + R_2(w,z)$.
\\
\indent
\textbf{b)} Using definition (\ref{R1}) of $R_1(w,z)$, we calculate 
$$
R_1(-z,z)=2\lambda_2\mu_1 z, \; \; \; R_1(z,z) = -2\mu_2z(2z-\lambda_1+\mu_1)
$$
so that, for given $z > 0$, $R_1(-z,z) > 0$ and $R_1(z,z) < 0$ since $\lambda_1 < \mu_1$. Further, the 3rd degree polynomial $R_1(\centerdot,z)$ has limits  $R_1(-\infty,z) = -\infty$ and $R_1(+\infty,z) = +\infty$. For given $z > 0$, real polynomial $R_1(\centerdot,z)$ has therefore 3 distinct real roots denoted by $\alpha_1(z)$, $\beta_1(z)$, $\gamma_1(z)$ which can be ordered as $\alpha_1(z) < -z < \beta_1(z) < z < \gamma_1(z)$. 
\\
\indent
By definition (\ref{R2}) of $R_2$, we similarly have
$$
R_2(-z,z) = 2\mu_1z(2z-\lambda_2+\mu_2), \; \; \; R_2(z,z) = -2\lambda_1\mu_2 z;
$$
identical arguments apply to show that, for given $z > 0$, real polynomial $R_2(\centerdot,z)$ has 3 distinct real roots denoted by $\alpha_2(z), \beta_2(z), \gamma_2(z)$ which can be ordered as $\alpha_2(z) < -z < \beta_2(z) < z < \gamma_2(z)$.

\subsection{Proof of Lemma \ref{iteration0}}
\label{A6}


Given $z > 0$, condition (\ref{intersect}) is equivalent to the existence of $z_1^* = B_1^{-1} \circ A_1(z)$, which is ensured provided that the inverse mapping $B_1^{-1}$ is locally defined. The existence of $B_1^{-1}$ follows from the local inversion theorem, claiming that $B_1$ is locally invertible at any point where its derivative is non-zero. In fact, definition (\ref{coord1}) implies that $B_1'(z) = 1 - \beta_1'(z)$; differentiating relation $R_1(\beta_1,z) = 0$ with respect to $z$ further gives
$$
\beta_1'(z) = - \frac{\partial_{z}R_1(\beta_1,z)}{\partial_{w}R_1(\beta_1,z)}
$$
(the derivative $\partial_{w}R_1(\beta_1,z)$ is non-zero since $w = \beta_1$ is a simple root of equation $R_1(w,z) = 0$) and $B_1'(z) = 0$ is equivalent to $\partial_{w}R_1(\beta_1,z) + \partial_{z}R_1(\beta_1,z) = 0$. As introduced in Appendix \ref{A3}.a, write then $R_1(w,z) = r_1(w+z,-w+z)$ where $r_1(s_1,s_2) = -(s_1+\mu_1)(s_2+\mu_2)(s_1 - K(s_1,s_2))$, so that
$$
\frac{\partial R_1}{\partial w} + \frac{\partial R_1}{\partial z} = 2 \frac{\partial r_1}{\partial s_1} = - \frac{2\lambda_2\mu_2 \left [ (s_1+\mu_1)^2 - \lambda_1\mu_1 \right ] }{s_1^2 - (\lambda-\mu_1)s_1 - \lambda_2\mu_1} 
$$
after using expression (\ref{T1T2}) for $s_2 = T_1(s_1)$. The latter is thus non-zero for any $s_1 \neq a_1^\pm$, where $a_1^\pm = -\mu_1 \pm \sqrt{\lambda_1\mu_1} < 0$ corresponds to the extremal points of function $T_1$ introduced in (\ref{statpoints1}); but by Lemma \ref{Roots}.b, abcissa $s_1 = \beta_1 + z$ is always positive for $z > 0$ and cannot therefore equal such negative values. The function $z > 0 \mapsto B_1(z)$ with positive derivative is therefore strictly increasing and there consequently exists a unique $z_1^* > 0$ such that $B_1(z_1^*) = A_1(z)$.
\\
\indent
We now show that $z_1^* > z$. Using definition (\ref{T1T2}) of function $T_1$ and the fact that $A_1 = T_1(a_1)$ and $B_1^* = T_1(b_1^*)$, condition (\ref{intersect}) simply reads
$$
\frac{a_1(a_1+\mu_1-\lambda_1)}{a_1^2 - (\lambda-\mu_1)a_1-\lambda_2\mu_1} = \frac{b_1^*(b_1^* + \mu_1-\lambda_1)}{(b_1^*)^2 - (\lambda-\mu_1)b_1^* - \lambda_2\mu_1}
$$
which, using $b_1^* \neq a_1$ and some simple algebra, reduces to relation (\ref{condition}). Parameter $2z_1^* = B_1^* + b_1^* = A_1 + b_1^*$ therefore equals
$$
2z_1^* = A_1 - \mu_1 \frac{a_1+\mu_1-\lambda_1}{a_1+\mu_1} = 2z - (a_1+\mu_1) + \frac{\lambda_1\mu_1}{a_1 + \mu_1}
$$
as claimed in (\ref{z2z2*}), after using the fact that $a_1 + A_1 = 2z$. The real point $\mathfrak{A}_1$ is on the segment of cubic $\mathcal{R}_1$ corresponding to $s_1 \in \; ]\sigma_1^-, a_1^+[$, that is, with abcissae $s_1$ larger than the asymptote $s_1 = \sigma_1^-$ and smaller than the horizontal tangent at $s_1 = a_1^+$; as $-\mu_1 < \sigma_1^-$, $\sigma_1^- < a_1 < a_1^+[$ implies $-\mu_1 < a_1 < a_1^+$ and it follows from the above expression of $z_1^*$ that $z_1^* > z$.


\subsection{Proof of Theorem \ref{Delta12}}
\label{A3bis}


We here detail the proof for item \textbf{a)} of Theorem \ref{Delta12}. 
\\
\indent
Write $R_j(w,z) = w^3 + R_{j1}(z)w^2 + R_{j2}(z)w + R_{j3}(z)$ with coefficients $R_{j\ell}(z)$, $1 \leq \ell \leq 3$, defined in (\ref{R1})-(\ref{R2}). By variable change $w = y - R_{j1}(z)/3$, $R_j(w,z)$ reduces to $\widetilde{R}_j(y,z) = y^3 + \widetilde{P}_jy + \widetilde{Q}_j$  with
\begin{equation}
\widetilde{P}_j = R_{j2} - \frac{R_{j1}^2}{3}, \; \; \widetilde{Q}_j = R_{j3} - \frac{R_{j1}R_{j2}}{3} +  \frac{2R_{j1}^3}{27};
\label{Cardanter}
\end{equation}
the discriminant of cubic polynomial $R_j(w,z)$ with respect to variable $w$ [\textbf{Cox}, p.16] is then given by
$$
\Delta_j = - (4\widetilde{P}_j^3 + 27\widetilde{Q}_j^2). 
$$
Using that general expression, the calculation of $\Delta_j(z)$ in terms of $z$ gives the polynomial
$$
\Delta_j(z) = \sum_{0 \leq \ell \leq 4} C_{j\ell} z^{4-\ell}, \; \; j \in \{1,2\},
$$
with degree $\leq 4$ and where real coefficients $C_{j\ell}$, $0 \leq \ell \leq 5$, are homogeneous polynomials of parameters $\lambda_1, \lambda_2, \mu_1, \mu_2$ with degree $\ell + 2$. In particular, the coefficient $C_{j0} = 16 \left [ (\mu_j-\lambda_j)^2 + \lambda_{3-j}^2 + 2\lambda_{3-j}(\lambda_j + \mu_j) \right ]$ of monomial $z^4$ in $\Delta_j(z)$ is non-zero, so that $\Delta_j(z)$ has exactly degree 4.
\\
\indent
Besides, using Cardano's formulae for the cubic equation [\textbf{Cox}, p.16], each solution $\epsilon_j(z) \in \{\alpha_j(z),\beta_j(z),\gamma_j(z)\}$ of equation $R_j(w,z) = 0$ can be written as 
\begin{equation}
\epsilon_j = -\frac{R_{j1}}{3} + \kappa^m \sqrt[3]{\frac{1}{2} \left ( - \widetilde{Q}_j + \sqrt{- \frac{\Delta_j}{27} } \right )} + \kappa^n \sqrt[3]{\frac{1}{2} \left ( - \widetilde{Q}_j - \sqrt{- \frac{\Delta_j}{27} } \right )}
\label{Cardan}
\end{equation}
where $\kappa = e^{2i\pi/3}$, the pair $(m,n)$ can take either value $(0,0)$, $(1,2)$ or $(2,1)$, and with real polynomials $\widetilde{Q}_j$ and $\Delta_j$ defined as above in (\ref{Cardanter}). 
\\
\indent
\textbf{I.} We first address the separation of the zeros of discriminant $\Delta_1(z)$. Since
$$
R_1(z,w) = 0 \Leftrightarrow K_1(z+w,z-w) = 0
$$
by (\ref{abexp}), Proposition \ref{Genus0} ensures that equation $R_1(w,z) = 0$ also represents algebraic curve $\mathcal{R}_1$ in $\widehat{\mathbb{C}} \times \widehat{\mathbb{C}}$ with genus $g = 0$. Now, recall [\textbf{Jon}, Theorem 4.16.3] that the genus $g$ of an irreducible algebraic curve over $\widehat{\mathbb{C}} \times \widehat{\mathbb{C}}$ with degree $d$ in variable $w$ and $M$ distinct ramification points with respective order $p_m \in \mathbb{N}$, $1 \leq m \leq M$, is given by Riemann-Hurwitz formula
\begin{equation}
g = \frac{p}{2} - d + 1
\label{RiemHurw}
\end{equation}
where $p = p_1 + ... + p_M$; such ramification points are
\begin{itemize}
\item [\textbf{I.A}] either a finite root $\eta$ of discriminant $\Delta_1(z) = 0$. By Cardano's formula (\ref{Cardan}), the ramification order of such a root $z$ for curve $R_1 = 0$ equals the ramification order of function $\sqrt{\Delta_1}$ at point $z$;
\item [\textbf{I.B}] or an algebraic  ingularity at $w = \infty$ and/or $z = \infty$.
\end{itemize}
In the present case, we have $g = 0$ and $d = 3$ for curve $R_1(w,z) = 0$, so we must have $p = 4$ by (\ref{RiemHurw}). 

Cases \textbf{I.A} and \textbf{I.B} above then specify as follows:
\\
\indent
- Case \textbf{I.A:} let $\eta^{(m)}$, $1 \leq m \leq 4$, denote the roots \footnote{Roots $\eta^{(m)}$, $1 \leq m \leq 4$, should also bear subscript "1" as they are related to $\Delta_1(z)$; we here omit this subscript to alleviate notation but mention it for completeness in next Subsection II of the present proof.} of discriminant $\Delta_1(z)$ with respective multiplicity order $k_m$. Then the ramification order $p_m$ of $\eta^{(m)}$ is that of function $\sqrt{\Delta_1(z)} \varpropto (z-\eta^{(m)})^{k_m/2}$, that is,
$$
p_m = 
\left\{
\begin{array}{ll}
0 \; \; \mathrm{if}  \; \;  k_m  \; \; \mathrm{is \; even}
\\ \\
1 \; \; \mathrm{if}  \; \;  k_m  \; \; \mathrm{is \; odd}.
\end{array} \right.
$$
In the present case, discriminant $\Delta_1(z)$ with degree 4 equals either
\\ \indent
\textit{i1)} $(z - \eta^{(1)})^4$, 
\\ \indent
\textit{i2)} $(z-\eta^{(1)})^3(z-\eta^{(2)})$ with $\eta^{(1)} \neq \eta^{(2)}$,
\\ \indent
\textit{i3)} $(z-\eta^{(1)})^2(z-\eta^{(2)})^2$ with $\eta^{(1)} \neq \eta^{(2)}$, 
\\ \indent
\textit{i4)}  $(z-\eta^{(1)})^2(z-\eta^{(2)})(z-\eta^{(3)})$ with distinct  $\eta^{(1)}, \eta^{(2)}, \eta^{(3)}$,
\\ \indent
\textit{i5)} $(z-\eta^{(1)})(z-\eta^{(2)})(z-\eta^{(3)})(z-\eta^{(4)})$ with distinct $\eta^{(1)}, \eta^{(2)}, \eta^{(3)}, \eta^{(4)}$,
\\
up to some multiplying constant. Using the above arguments, the total ramification order $p$ associated with each subcase \textit{i1)-i5)} is then tabulated as follows:
$$
\begin{array}{|l|c|c|c|}
\hline \mathrm{Distinct \; roots \; \; } \eta^{m}, 1 \leq m \leq 4 &  \mathrm{Order} \; p_m & \mathrm{Total \; order \; } p = p_1 + ... + p_4 \\[1 mm] 
\hline 
i1) \;  \; \eta^{(1)} \;                          & 0                   & p = 0 \\[1 mm]
i2) \;  \; \eta^{(1)}, \eta^{(2)} \;              & 1, 1                & p = 2 \\[1 mm]
i3) \;  \; \eta^{(1)}, \eta^{(2)} \;              & 0, 0                & p = 0 \\[1 mm]
i4) \;  \; \eta^{(1)}, \eta^{(2)}, \eta^{(3)} \;  & 0, 1, 1         & p = 2 \\[1 mm]
i5) \;  \; \eta^{(1)}, \eta^{(2)}, \eta^{(3)}, \eta^{(4)} \; & 1, 1, 1, 1 &  \mathbf{p = 4} \\[1 mm]
\hline
\end{array}
$$
\indent
- Case \textbf{I.B:} as to possible ramification points at infinity, we easily verify that equation $\zeta_1^3 R_1(1/\zeta_1,z) = 0$ has no solution for $\zeta_1 = 0$ and finite $z$; similarly, equation $\zeta_2^3 R_1(w,1/\zeta_2) = 0$ has no solution for finite $w$ and $\zeta_2 = 0$. There are consequently no ramification points are either $w = \infty$ or $z = \infty$. Finally, we verify that equation $\zeta_1^3 \zeta_2^3 R_1(1/\zeta_1,1/\zeta_2) = 0$ has the solution $(0,0)$ and locally defines a unique branch $\zeta_2 = O(\zeta_1)$; $(w = \infty, z = \infty)$ is consequently not a ramification point.
\\
\indent
As a conclusion, the only case giving a total ramification order $p = 4$ corresponds to subcase \textit{i5)} for which discriminant $\Delta_1$ has four distinct roots. A similar statement symmetrically holds for discriminant $\Delta_2$. 
\\
\indent
\textbf{II.} We finally investigate the number of real roots of discriminant $\Delta_1(z)$. Any root $z$ of $\Delta_1(z)$ is such that polynomial equations 
$$
R_1(w,z) = \frac{\partial R_1}{\partial w}(w,z) = 0
$$
have a common solution in $w$. Writing $R_1(w,z) = r_1(s_1,s_2)$ as in \textbf{�\ref{A3}.a}, the second condition reads
\begin{equation}
0 = \frac{\partial R_1}{\partial w} = \frac{\partial r_1}{\partial s_1} \cdot \frac{\partial s_1}{\partial w} + \frac{\partial r_1}{\partial s_2} \cdot \frac{\partial s_2}{\partial w} = \frac{\partial r_1}{\partial s_1} - \frac{\partial r_1}{\partial s_2}
\label{impl1}
\end{equation}
by variable change (\ref{varch}) and the chain rule; on the other hand, $r_1(s_1,s_2) = 0$ is equivalent to $s_2 = T_1(s_1)$ by (\ref{K1z1z2}) and relation $r_1(s_1,T_1(s_1)) = 0$ implies in turn
$$
\frac{\mathrm{d}T_1}{\mathrm{d}s_1}(s_1) = - \frac{\partial r_1}{\partial s_1}(s_1,T_1(s_1)) \left ( \frac{\partial r_1}{\partial s_2} (s_1,T_1(s_1)) \right )^{-1} = -1
$$
by the Implicit function Theorem and relation (\ref{impl1}). We conclude that a root $z$ of $\Delta_1(z)$ corresponds to a point $(s_1,T_1(s_1))$ in the $(O,S_1,S_2)$ frame such that $T_1'(s_1) = -1$; equivalently, the line $S_1 + S_2 = 2z$ is tangent to $\mathcal{R}_1$ at $(s_1,T_1(s_1))$.
\\
\indent
Now, compute
\begin{equation}
\frac{\mathrm{d}T_1}{\mathrm{d}s_1}(s_1) = \frac{\lambda_2\mu_2}{d_1(s_1)^2}\left ( (s_1 + \mu_1)^2 - \lambda_1\mu_1 \right )
\label{T1prime}
\end{equation}
after definition (\ref{T1T2}) of $T_1(s_1)$, with denominator $d_1(s_1) = s_1^2 - (\lambda - \mu_1)s_1 - \lambda_2\mu_1$. Recall that the poles of $T_1$ are the zeros $\sigma_1^-$ and $\sigma_1^+$ of $d_1(s_1)$, as defined in (\ref{poles1s2}); calculating $d_1(0) = - \lambda_2\mu_1 < 0$, $d_1(a_1^+) = - \sqrt{\lambda_1\mu_1}(\lambda + \mu_1 - 2 \sqrt{\lambda_1\mu_1}) < 0$ and $d_1(a_1^-) = \sqrt{\lambda_1\mu_1}(\lambda + \mu_1 + 2 \sqrt{\lambda_1\mu_1}) > 0$ implies  inequalities $a_1^- < \sigma_1^- < a_1^+ < 0 < \sigma_1^+$. The sign of $(s_1 + \mu_1)^2 - \lambda_1\mu_1$ in (\ref{T1prime}) then enables us to derive that
\begin{itemize}
\item[\textbf{II.1}] $T_1^{'}(s_1) < 0$ for all $s_1 \in \; ]a_1^-, \sigma_1^-[ \; \cup \; ]\sigma_1^-,a_1^+[$ with specifc values $T_1^{'}(a_1^-) = 0$, $T_1^{'}(\sigma_1^- - 0) = -\infty$, $T_1^{'}(\sigma_1^-+0) = -\infty$ and $T_1^{'}(a_1^+) = 0$;
\item[\textbf{II.2}] $T_1^{'}(s_1) > 0$ otherwise.
\end{itemize}
It then follows from \textbf{II.1} that equation $T_1^{'}(s_1) = -1$ has at least two distinct real roots $a_1^{(2)}$, $a_1^{(1)}$ such that $a_1^- < a_1^{(2)} < \sigma_1^-$ and $\sigma_1^- < a_1^{(1)} < a_1^+$. Besides, using (\ref{T1prime}), we calculate
$$
\frac{\mathrm{d}^2 T_1}{\mathrm{d}s_1^2}(s_1) = -\frac{2\lambda_2\mu_2}{d_1(s_1)^3}C_1(s_1)
$$
where $C_1(s_1) = s_1^3 + 3\mu_1 s_1^2 + 3\mu_1(\mu_1-\lambda_1)s_1 + \mu_1((\mu_1-\lambda_1)^2 + \lambda_1\lambda_2)$. 

Cubic polynomial $C_1(s_1)$ has real coefficients along with a negative discriminant $-27\lambda_1^2\mu_1^2((\mu_1-\lambda_1)^2 + \lambda_2^2 + 2\lambda_2(\lambda_1 + \mu_1))$; it has therefore [\textbf{Cox04}, Theorem 1.3.1] a unique real root $s_1^*$, corresponding to a unique real extremum for first derivative $T_1'$. By properties \textbf{II.1} and \textbf{II.2} above, $T_1'$ is positive on interval $]-\infty,a_1^-[$ with $T_1'(-\infty) = T_1'(a_1^-) = 0$; it follows that extremum $s_1^*$ verifies $s_1^* < a_1^-$. Correspondingly, function $T_1'$ is monotonous on either interval $]a_1^-,\sigma_1^-[$ or $]\sigma_1^-,a_1^+[$; the distinct roots $a_1^{(2)}$, $a_1^{(1)}$ determined above are therefore the unique real roots of equation $T_1'(s_1) = -1$. Correspondingly, the discriminant $\Delta_1(z)$ has only two real roots $\eta_1^{(2)} < \eta_1^{(1)} < 0$ with
$$
\eta_1^{(2)} = \frac{a_1^{(2)} + T_1(a_1^{(2)})}{2}, \; \; \; \eta_1^{(1)} = \frac{a_1^{(1)} + T_1(a_1^{(1)})}{2}.
$$ 
As polynomial $\Delta_1(z)$ is real, its non real roots $\eta_1^{(3)}$ and $\eta_1^{(4)}$ are complex conjugates. Similar conclusions hold for discriminant $\Delta_2(z)$.
\begin{remark}
From the above discussion, the positivity of second order derivative $T_1''(s_1)$ for $s_1 \in \; ]\; \sigma_1^-,\sigma_1^+ \; [$ implies that function $T_1$ (and equivalently, curve $\mathcal{R}_1$) is convex on that interval. Mutatis mutandis, function $T_2$ (and equivalently, curve $\mathcal{R}_2$) is convex on interval $]\sigma_2^-,\sigma_2^+ \; [$.
\label{Rem2}
\end{remark}

\subsection{Possible vanishing of denominator $E(z)$}
\label{Psing}


The denominator $E(z)$ in (\ref{denE}) vanishes if and only if
\begin{align}
E_0(z) & = (\mu_1+\xi_1^-(s_2))(\mu_2+\xi_2^-(s_1)) - (\mu_1+s_1)(\mu_2+s_2)
\nonumber \\
& = \mu_1(\xi_2^-(s_1)-s_2) + \mu_2(\xi_1^-(s_2) - s_1) + \xi_1^-(s_2)\xi_2^-(s_1) - s_1s_2
\nonumber \\
& = (\alpha_1(z)-\gamma_2(z))(2z + \mu_1+\mu_2) = 0
\nonumber
\end{align}
after using identities $s_1 = z + \gamma_2(z)$, $s_2 = z - \alpha_1(z)$, $\xi_2^-(s_1) = z - \gamma_2(z)$ and $\xi_1^-(s_2) = z + \alpha_1(z)$; it follows that $E_0(z) = 0$ if and only if
\begin{equation}
\alpha_1(z) = \gamma_2(z) \; \; \; \mathrm{or} \; \; \; z = - \frac{\mu_1+\mu_2}{2}.
\label{conditionAC0}
\end{equation}
As mentioned in the course of the proof of Proposition \ref{extendM1M2}, we now verify the following.
\begin{lemma}
Denominator $E(z)$ does not vanish for $\Re(z) > \max(\eta_1^{(1)},\eta_2^{(1)})$.
\label{lemmaE}
\end{lemma}
\begin{proof} In view of the latter calculation, the assertion is verified if we show  that each condition (\ref{conditionAC0}) cannot correspond to a zero of $E_0(z)$ for $\Re(z) > \max(\eta_1^{(1)},\eta_2^{(1)})$.
\\
\indent
\textbf{1)} First condition $\alpha_1(z) = \gamma_2(z)$ in (\ref{conditionAC0}) implies that polynomials $R_1(w,z)$ and $R_2(w,z)$ with unknown $w$ have a common root; their resultant $R(z)$ must consequently vanish. Using definitions (\ref{R1}) and (\ref{R2}) for $R_1(w,z)$ and $R_2(w,z)$, we calculate $R(z) = 8\lambda_1\lambda_2\mu_1\mu_2zP(z)(2z+\mu_1+\mu_2)^2$ with polynomial $P(z)$ defined in (\ref{PolP}), hence
$$
R(z) = 0 \Leftrightarrow z = 0, P(z) = 0 \; \mathrm{or} \; z = -(\mu_1+\mu_2)/2.
$$
We first easily verify that $z = 0$ or any root $z$ of $P(z)$ corresponds to a common root $\beta_1(z) = \beta_2(z)$ to $R_1(w,z)$ and $R_2(w,z)$, and therefore not to $\alpha_1(z) = \gamma_2(z)$. As to $z = -(\mu_1+\mu_2)/2$, note that if $R_1(w,z) = R_2(w,z) = 0$ for some $z$, then identity (\ref{R1R2}) implies either $w = 0$ (which again gives $zP(z) = 0$, but this was excluded above), $w = -\mu_1 - z$ or $w = \mu_2 + z$; for $z = -(\mu_1+\mu_2)/2$, we then have $w = (\mu_2-\mu_1)/2$ and it is easily verified that $R_1(w,z) \neq 0$ and $R_2(w,z) \neq 0$ for such specific values of $w$ and $z$. The above discussion therefore implies that equation $\alpha_1(z) = \gamma_2(z)$ has no solution. 
\\
\indent
\textbf{2)} Now turn to the possible zero $z = -(\mu_1+\mu_2)/2$ of $E(z)$, as expressed by 2nd condition (\ref{conditionAC0}). Following Theorem \ref{Delta12}.a, discriminant  $\Delta_1(z)$ has only two real zeros $\eta_1^{(2)} < \eta_1^{(1)} < 0$; by Appendix \ref{A3bis}, the highest degree monomial $C_{10}z^4$ of $\Delta_1(z)$ is positive so that $\Delta_1(-\infty) = \Delta_1(+\infty) = +\infty$. We thus deduce that  $\Delta_1(z) > 0$ if and only if $z < \eta_1^{(2)}$ or $z > \eta_1^{(1)}$. Now,
$$
\Delta_1\left ( - \frac{\mu_1+\mu_2}{2} \right) = (\lambda_1\mu_1-\lambda_2\mu_2)^2 \left[(\lambda_1-\mu_1)^2 + \lambda_2^2 + 2\lambda_1\lambda_2 + 2\lambda_2(\mu_1 + 2\mu_2) \right]
$$
is positive, whereas
\begin{align}
\Delta_1'\left ( - \frac{\mu_1+\mu_2}{2} \right) = & -4(\lambda_1+\lambda_2+\mu_2) \Big [ \lambda_1\mu_1(\lambda_1-\mu_1)^2 + \lambda_2\mu_2((\lambda_2+\mu_1)^2+5\lambda_2\mu_2) 
\nonumber \\
& + \lambda_1^2\lambda_2(2\mu_1+\mu_2) + \lambda_1\lambda_2\mu_1(2\mu_1 + \mu_2) + \lambda_1\lambda_2^2(\mu_1+2\mu_2) \Big ]
\nonumber
\end{align}
is negative. As a third degree polynomial of variable $z$, derivative $\Delta_1'(z)$ can have
\begin{itemize}
\item[\textbf{2.a)}] either a single real zero $\theta_1^{(0)}$: by the above observations, $\Delta_1$ has then a unique minimum at $\theta_1^{(0)} \in \; ]\eta_1^{(2)},\eta_1^{(1)}[$ and the signs of $\Delta_1(-(\mu_1+\mu_2)/2)$ and $\Delta_1'(-(\mu_1+\mu_2)/2)$ calculated above imply that $-(\mu_1+\mu_2)/2 < \eta_1^{(2)}$;
\item[\textbf{2.b)}] or three distinct real zeros, say, $\theta_1^{(3)} < \theta_1^{(2)} < \theta_1^{(1)}$: 

- first assume that $\theta_1^{(3)} < \theta_1^{(2)} < \eta_1^{(2)} < \theta_1^{(1)} < \eta_1^{(1)}$; $\Delta_1(z)$ and $\Delta_1'(z)$ are respectively positive and negative for $z < \eta_1^{(2)}$ only; we thus again conclude that $-(\mu_1+\mu_2)/2 < \eta_1^{(2)}$; 

- on the contrary, assume that $\eta_1^{(2)} < \theta_1^{(3)} < \eta_1^{(1)} < \theta_2^{(2)} < \theta_1^{(1)}$; inflexion points $\zeta_1'' < \zeta_1'$ of $\Delta_1$ are then such that $\Delta_1''(z) < 0$ for $z \in ]\zeta_1'',\zeta_1'[$ and $\Delta_1''(z) > 0$ otherwise. We first note that $\Delta_1'''(z)$ is a linear and increasing function of $z$, vanishing inside interval $]\zeta_1'',\zeta_1'[$; as 
$$
\Delta_1'''\left ( - \frac{\mu_1+\mu_2}{2} \right) = -96(\lambda_1+\lambda_2+\mu_1) \left[ (\lambda_1-\mu_1)^2 + \lambda_2^2 + 2\lambda_1\lambda_2 + \lambda_2(2\mu_1+\mu_2) \right ]
$$
is negative, we thus have $-(\mu_1+\mu_2)/2 < \zeta_1'$. Besides, we calculate
\begin{align}
\frac{1}{8}\Delta_1''\left ( - \frac{\mu_1+\mu_2}{2} \right) = & \; \mathring{\lambda}_1^4 + \lambda_2^4 + \mathring{\mu}_1^4 + 2\lambda_1^3(2\lambda_2+\mathring{\mu}_1) + 4\lambda_2^3(\mu_1+2\mu_2) \; + 
\nonumber \\
& 4\lambda_2\mu_2^2(\mu_1+2\ddot{\mu}_2) + \lambda_2^2(6\mu_1^2+16\mu_1\mu_2+\mu_2^2) \; +
\nonumber \\
& \lambda_1^2(6\lambda_2^2-6\mathring{\mu}_1^2+8\lambda_2(\mu_1+\mu_2)) \; +
\nonumber \\
& 2\lambda_1(2\lambda_2^3+\mathring{\mu}_1^3+2\lambda_2\mu_1(2\mu_1-\ddot{\mu}_2)+\lambda_2^2(5\mu_1+8\mu_2));
\nonumber
\end{align}
grouping ringed terms in the expression above gives
$$
\lambda_1^4+\mu_1^4+2\lambda_1^3\mu_1-6\lambda_1\mu_1^2+2\lambda_1\mu_1^3 = \lambda_1^4 \left[ \frac{\mu_1}{\lambda_1} - 1 \right ]^2 \left[ \left( \frac{\mu_1}{\lambda_1} \right)^2 + 4 \frac{\mu_1}{\lambda_1} + 1 \right ] > 0;
$$
similarly, grouping twice dotted terms implies
\begin{align}
& 4\lambda_2\mu_2^2 \times 2\ddot{\mu}_2 - 2\lambda_1 \times 2\lambda_2\mu_1\ddot{\mu}_2 = 
\nonumber \\
& 8\lambda_2\mu_1^2\mu_2 - 4\lambda_1\lambda_2\mu_1\mu_2 >  8\lambda_1\lambda_2\mu_1\mu_2 - 4\lambda_1\lambda_2\mu_1\mu_2 > 0
\nonumber
\end{align}
since $\lambda_1 < \mu_1$. We thus conclude from the above calculations that the second order derivative $\Delta_1''(-(\mu_1+\mu_2)/2)$ is positive, and the arguments above imply that $-(\mu_1+\mu_2)/2 < \eta_1^{(2)}$.
\end{itemize}
We thus conclude from previous steps \textbf{1)} and \textbf{2)} that $E_0(z)$, and therefore $E(z)$, does not vanish for $\Re(z) > \eta_1^{(1)}$; similar calculations show that $E(z)$ does not vanish for $\Re(z) > \eta_2^{(1)}$. The expected conclusion follows. \end{proof}


\section{References}


\textbf{[Bel09]} M.C.Beltrametti, E.Carletti, D.Gallarati, G.M.Bragadni, \textit{Lectures on curves, surfaces and projective varieties, A classical view of algebraic geometry}, ed. European Mathematical Society, Textbooks in Mathematics, 2009
\\
\indent
\textbf{[Cox04]} D.A.Cox, \textit{Galois Theory}, ed. J.Wiley Interscience, 2004
\\
\indent
\textbf{[Doe58]} G.Doetsch, \textit{Einf�hrung in Theorie und Anwendung der Laplace Transformation}, ed. Birkh�user, 1958
\\
\indent
\textbf{[Fis01]} G.Fisher, \textit{Plane Algebraic Curves}, ed. American Mathematical Society, 2001
\\
\indent
\textbf{[Gui12]} F.Guillemin, A.Simonian, \textit{Stationary analysis of the SQF service policy}, Submitted for publication, 2012
\\
\indent
\textbf{[Mil06]} J.W.Milnor, \textit{Dynamics in one complex variable}, 3rd edition, Princeton University Press, 2006


\end{document}